\documentclass[a4paper,superscriptaddress,floatfix,nofootinbib,onecolumn,longbibliography, notitlepage]{revtex4-1}
\usepackage{bbm}
\usepackage{bbold}
\usepackage{bbm}
\usepackage[pdftex]{graphicx}
\usepackage{latexsym,amsmath,verbatim,amssymb,txfonts}
\usepackage{lipsum}
\usepackage{color}
\usepackage{rotating}
\usepackage{verbatim}
\usepackage{physics}
\usepackage{braket}
\usepackage{multirow}
\usepackage{hyperref}
\usepackage[english]{babel}
\usepackage{comment}
\usepackage{bm}
\usepackage{amsmath}
\usepackage{amsfonts}
\usepackage{amssymb}
\usepackage{float}
\usepackage{hyperref}
\usepackage{color}
\usepackage{braket}
\usepackage{ulem}
\usepackage{blkarray, bigstrut}

\renewcommand\bra[1]{{\langle{#1}|}}
\renewcommand\ket[1]{{|{#1}\rangle}}

\definecolor{internationalkleinblue}{rgb}{0.0, 0.18, 0.65}
\definecolor{brickred}{rgb}{0.8, 0.25, 0.33}
\definecolor{darkviolet}{rgb}{0.58, 0.0, 0.83}

\begin{document}
\title{Revisiting weak values through non-normality}
\author{Lorena Ballesteros Ferraz}
\email{lorena.ballesteros@unamur.be}
\affiliation{naXys, Namur Institute for Complex Systems, University of Namur, Belgium}
\affiliation{NISM, Namur Institute of Structured Matter, University of Namur, Belgium}
\affiliation{Research Unit Lasers and Spectroscopies (UR-LLS), Department of Physics, University of Namur, Belgium}

\author{Riccardo Muolo}
\affiliation{Department of Systems and Control Engineering, Tokyo Institute of Technology, 2 Chome-12-1 Ookayama, Tokyo 152-8550, Japan}
\affiliation{naXys, Namur Institute for Complex Systems, University of Namur, Belgium}
\affiliation{Department of Mathematics, University of Namur, Belgium}

\author{Yves Caudano}
\affiliation{naXys, Namur Institute for Complex Systems, University of Namur, Belgium}
\affiliation{NISM, Namur Institute of Structured Matter, University of Namur, Belgium}
\affiliation{Research Unit Lasers and Spectroscopies (UR-LLS), Department of Physics, University of Namur, Belgium}

\author{Timoteo Carletti}
\affiliation{naXys, Namur Institute for Complex Systems, University of Namur, Belgium}
\affiliation{Department of Mathematics, University of Namur, Belgium}

\begin{abstract}
Quantum measurement is one of the most fascinating and discussed phenomena in quantum physics, due to the impact on the system of the measurement action  and the resulting interpretation issues. Scholars proposed weak measurements to amplify measured signals by exploiting a quantity called a weak value, but also to overcome philosophical difficulties related to the system perturbation induced by the measurement process. The method finds many applications and raises many philosophical questions as well, especially about the proper interpretation of the observations. In this paper, we show that any weak value can be expressed as the expectation value of a suitable non-normal operator. We propose a preliminary explanation of their anomalous and amplification behavior based on the theory of non-normal matrices and their link with non-normality: the weak value is different from an eigenvalue when the operator involved in the expectation value is non-normal. Our study paves the way for a deeper understanding of the measurement phenomenon, helps the design of experiments, and it is a call for collaboration to researchers in both fields to unravel new quantum phenomena induced by non-normality.
\end{abstract}

\maketitle

\section{Introduction}
\color{black}
Weak values play a role similar to expectation values in certain types of quantum measurements. However, they can be anomalous, meaning that they lie outside the range of eigenvalues of the measured operator, which represents a determined quantum property. Hence they can become complex numbers and be unbounded \cite{luo2017precision, dixon2009ultrasensitive, hallaji2017weak}. Anomalous weak values are used to amplify signals \cite{hosten2008observation}, to measure complex properties such as wave functions \cite{lundeen2011direct} or expectation values of non-Hermitian operators \cite{pati2015measuring}, and to study fundamental quantum phenomena, like paradoxes \cite{aharonov2002revisiting, lundeen2009experimental, matzkin2013three, gul2022implementation}.  Anomalous weak values evidence an intrinsic property of non-classicality, called contextuality \cite{pusey2014anomalous, kunjwal2019anomalous}. A particular class of anomalous weak values is associated to amplification. This phenomenon occurs when the modulus of the weak value is larger than the modulus of all the observable eigenvalues. Such weak values are said to be in the amplifying range.

\begin{figure}[b]
 \centering
\includegraphics[width=0.8\textwidth]{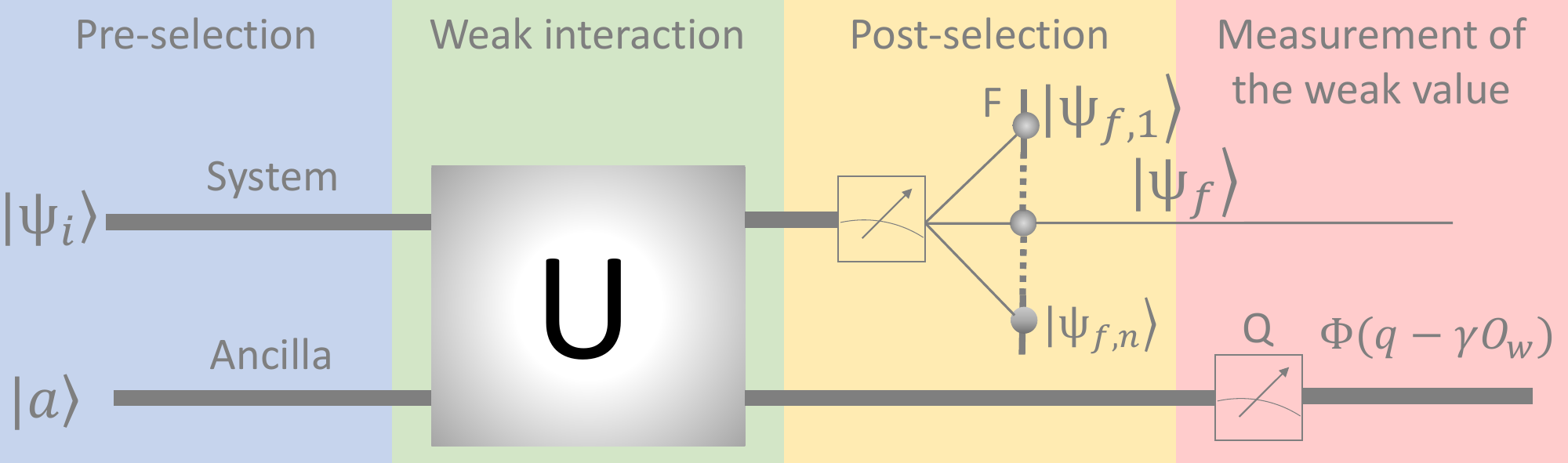}
\caption{Scheme of a weak measurement with post-selection. In the first step, pre-selection, the initial state of the system is set to $\ket{\psi_i}$ (blue). Then, the system and ancilla interact through a unitary operator $U=e^{-i\gamma\hat{O}\otimes\hat{P}}$ (green). This interaction should be weak, meaning that the interaction strength $\gamma$ should be small. After the weak interaction, a projective measurement is executed in the system and the final state is chosen to be $\ket{\psi_f}$ (yellow). If the post-selection is successful, the ancilla position is measured by applying a projective measurement in the ancilla (pink). The expectation value of the position is shifted by a quantity that is proportional to the weak value $O_w$ of the system's observable $\hat{O}$ for the pre- and post-selected system states. \label{fig:scheme_weak_measurement}}
\end{figure}

Weak values were defined in the context of the weak von Neumann measurement protocol with post-selection \cite{aharonov1988result}. This procedure, schematically shown in Fig.~\ref{fig:scheme_weak_measurement}, is called a weak measurement. The latter is performed by using the system under scrutiny and a meter (or ancilla), each one represented by a horizontal line in the scheme depicted in Fig.~\ref{fig:scheme_weak_measurement}. The measurement protocol proceeds in four steps. Firstly, the pre-selection step chooses the system initial state $\ket{\psi_i}$ (see the blue rectangle in Fig.~\ref{fig:scheme_weak_measurement}). Similarly, the known state $\ket{\alpha}$ defines the initial reference point of the ancilla. In practice, these states are typically obtained via projective measurements. Secondly, a reversible system--ancilla interaction is implemented through a unitary operator $\hat{U}=e^{i\gamma\hat{O}\otimes\hat{P}}$ (green panel in Fig.~\ref{fig:scheme_weak_measurement}). The observable of interest $\hat{O}$ operates onto the system space, while the momentum operator $\hat{P}$, called the pointer of the protocol, belongs to the meter space; $\gamma$ is the interaction strength (for simplicity, we set the normalized Plank constant $\hbar=1$ in the unitary operator). This step entangles the system and the ancilla. For a weak measurement, the interaction strength should be small, enabling the expansion of the unitary operator in a Taylor series up to the first order. In this way, the measurement only slightly modifies the system state, limiting the measurement-induced perturbation. In return, the extracted information in a single measurement is very limited as well. The third step, post-selection, consists of a projective measurement imposed on the system, followed by filtering to set into a pre-determined final state $\ket{\psi_f}$  the system at the end of the protocol (see the yellow panel in Fig.~\ref{fig:scheme_weak_measurement}). Eventually, information about the pre- and post-selected system is extracted by reading out the ancilla wave function conditioned on successful post-selection (see the pink panel in Fig.~\ref{fig:scheme_weak_measurement}).

At the end of the protocol, the ancilla wave function is shifted in position representation by a quantity proportional to the real part of the weak value, $O_w=\frac{\bra{\psi_f}\hat{O}\ket{\psi_i}}{\braket{\psi_f|\psi_i}}$, where $\ket{\psi_i}$ and $\ket{\psi_f}$ are the pre- and post-selected states, while the ancilla wave function in momentum representation is shifted by a quantity proportional to the imaginary part of the weak value. Since the interaction strength is weak, many measurements must be performed and averaged to determine the weak value precisely. In the absence of post-selection, the von Neumann protocol describes standard measurements for arbitrary interaction strengths, where the measuring device --represented by the ancilla-- is treated quantumly. Thus, in quantum measurements involving weak interaction and post-selection, the weak value substitutes the expectation value of the observable in the average wave function shift of the meter. Nonetheless, certain authors posit that weak values exhibit a closer affinity to eigenvalues than to the expectation value \cite{vaidman2017weak}.

The significance of weak values in physics extends beyond weak measurements. They appear in strong quantum measurements \cite{de2022role}, and in homodyne measurements \cite{wiseman2002weak}. In general measurements, the real and the imaginary components of the weak value are linked to the optimal conditional estimate of the observable and the inaccuracy, respectively \cite{hall2004prior, dressel2015weak, dressel2012significance, hofmann2011uncertainty}. They also contribute intrinsically to dynamical phenomena \cite{dressel2015weak}. While the real and imaginary parts generate the meter shifts in typical weak measurements, the modulus and argument of weak values also hold significance \cite{mc2016,mc2017,lbf2022}, particularly as the argument characterizes geometric phases within the quantum state manifold. In this work, special attention will be paid to the modulus of weak values.

The weak value of a generic observable $\hat{O}$, can always be expressed as the expectation value of an operator, defined in terms of the pre-selected state, the observable and the post-selected state, $O_w=\text{Tr}\left[\hat{\Pi}_i\hat{A}\right]$, with $\hat{A}=\frac{\hat{O}\hat{\Pi}_i}{\text{Tr}\left[\hat{\Pi}_f\hat{\Pi}_i\right]}$ and $\hat{\Pi}_i=\ket{\psi_i}\bra{\psi_i}$, resp.  $\hat{\Pi}_f=\ket{\psi_f}\bra{\psi_f}$, the projector of the pre-selected, resp. post-selected, state. This operator is not necessarily Hermitian. Most non-Hermitian operators are also non-normal, meaning that $\hat{A}\hat{A}^\dag\neq \hat{A}^\dag\hat{A}$. The choice of definition of $\hat{A}$ is arbitrary to some extent, as the weak value could also be defined in terms of an operator involving the post-selected state $\hat{A}'$, $O_w=\text{Tr}\left[\hat{\Pi}_f\hat{A}'\right]$, with $\hat{A}'=\frac{\hat{\Pi}_f\hat{O}}{\text{Tr}\left[\hat{\Pi}_f\hat{\Pi}_i\right]}$. In the framework of matrices, relevant to this paper, the above definition of non-normality implies that the matrix cannot be diagonalized through an orthonormal transformation \cite{trefethen}. Non-normality is a stronger relation of asymmetry than simple non-Hermicity: in fact, all non-normal matrices are non-Hermitian, but there is a class of non-Hermitian matrices which are normal, the circulant matrices \cite{entropy}. Such stronger asymmetry can be thought of as a hierarchical structure of the matrix. This clearly emerges when we consider networks: in such context, the adjacency matrix represents the way in which the nodes (i.e, the units of the network) are connected to each other; when the structure of the connections is hierarchical and directed, the adjacency matrix is non-normal \cite{johnson_directed,asllani_leaders}. The latter framework results  particularly interesting for applications, as real-world networks are non-normal \cite{malbor_teo}.

The first main result of this work is to show that the operators $\hat{A}$ and $\hat{A}'$ must be non-normal for the weak value to become anomalous.  When imposing an arbitrary post-selection, the weakly measured quantity changes from an expectation value of a Hermitian operator representing the observable of interest, to an expectation value of a specific non-normal operator. Interpreting weak values from this point of view might involve studying the energetics of the protocol, especially the post-selection process. Possible links with open systems and non-Hermitian quantum physics \cite{moiseyev2011non} might appear, as post-selection involves a discarding process.

Non-normality of matrices and operators has been studied in numerical analysis by Trefethen and collaborators \cite{trefethen,toeplitz_tref}, triggered by the effects observed in fluid dynamics \cite{trefethen2}. In fact, such property makes the matrix more sensible to perturbations, which results in an amplifying effect of the latter, with dramatic consequences on the dynamics. For example, in fluid dynamics, it gives rise to a premature emergence of turbulence \cite{trefethen}, while in epidemics it may lead to a lower threshold for an outbreak \cite{top_resilience}. Effects of non-normality have also been studied in ecology \cite{neub_cas,murray_neub_cas}, Turing pattern formation \cite{jtb}, chemical reaction networks \cite{mott_nn}, and synchronization dynamics \cite{entropy}. Such a change in the behavior of the system can be ascribed to the possible emergence of an initial transient growth, whose intensity is proportional to the non-normality of the system \cite{malbor_teo}. The degree of non-normality is estimated by the spectral properties of the matrix, resulting in different metrics \cite{trefethen,malbor_teo}. Those metrics will be used in the following to assess the strength of the anomalous weak value. In particular, we will be interested in using the Henrici departure from normality. 

In this paper, we demonstrate that non-normality is a necessary condition for obtaining a weak value that is different from an eigenvalue of the observable. We will show that, in order to have a large weak value, the operator $\hat{A}$ and $\hat{A}'$ must be non-normal; hence, the latter property is necessary to obtain amplification. Furthermore, by comparing the modulus of the weak value and the Henrici departure from normality with varying pre- and post-selected states, we make a direct connection between the degree of weak-value amplification and the degree of non-normality of the operator $\hat{A}$. As a next step, we consider a varying observable, which is the other component involved in the weak value. Ordinarily, experimenters select pre- and post-selected states that are nearly orthogonal in order to achieve amplification. However, there are cases in which we have some freedom in the choice of the observable. In this study, we demonstrate the critical role that this choice plays in the amplification yield. By varying the observable, we discovered that the maximum weak value always occurs at the arithmetic average of the maximum Henrici departure from normality of both matrices $\hat{A}$ and $\hat{A}'$. Furthermore, the weak value tends toward infinity when the points at which the matrices $\hat{A}$ and $\hat{A}'$ are degenerate and nilpotent coincide.

The paper is organized as follows: in the next section, we formally express the weak value as the expectation value of a non-normal operator. In section \ref{sec:numerical_results}, we discuss our numerical results, showing the correlation between the non-normality of the operator and the amplification of the weak value. Then, in section \ref{new_expression}, we study the correlation between the Henrici departure from normality and the weak value when varying the analyzed observable, showing that the anomalous weak value is still correlated to non-normality. Finally, section \ref{sec:conclusions} is left for the discussion of the results and conclusions.
\section{Weak values as expectation values of non-normal operators}
The aim of this section is to show the importance of non-normality for the emergence of weak values. To achieve this goal, we rewrite the weak value as the expectation value of a non-normal operator, $\hat{A}$. Then, we prove that when the latter operator is normal, the weak value is simply an eigenvalue of the observable, $\hat{O}$. This shows that weak values are directly linked to non-normal operators.

Let us consider a system that is initialized on the pure quantum state $\ket{\psi_i}$. The projector representing this state is $\hat{\Pi}_i=\ket{\psi_i}\bra{\psi_i}$. The post-selected projector is $\hat{\Pi}_f=\ket{\psi_f}\bra{\psi_f}$ and the observable of interest, namely a Hermitian operator, is $\hat{O}$.

The weak value of the system can be expressed as
\begin{equation}
O_w=\frac{\bra{\psi_f}\hat{O}\ket{\psi_i}}{\braket{\psi_f|\psi_i}}=\frac{\text{Tr}\left[\hat{\Pi}_f\hat{O}\hat{\Pi}_i\right]}{\text{Tr}\left[\hat{\Pi}_f\hat{\Pi}_i\right]}\, .
\label{eq:weak_value_in_terms_of_trace}
\end{equation}
A simple algebraic manipulation of the previous formula expresses the weak value in terms of the expectation value of the operator $\hat{A}$, 
\begin{equation}
O_w=\bra{\psi_f}\hat{A}\ket{\psi_f}\, ,
\label{eq:weak_value_in_terms_of_A}
\end{equation}
where the non-normal operator $\hat{A}$ is defined in terms of the operator $\hat{O}$ and the pre- and post-selected states as follows:
\begin{equation}
\hat{A}=\frac{\hat{O}\hat{\Pi}_i}{\text{Tr}\left[\hat{\Pi}_f\hat{\Pi}_i\right]}\, .
\label{eq:definition_matrix_A}
\end{equation}

To measure the non-normality of a matrix $M$, we use the Henrici departure from normality, $d_f$, defined as
\begin{equation}
\label{eq:Henrici_departure_from_normality}
d_f\left(M\right)=\sqrt{||M||^2_F-\sum_{i=1}^{n}|\lambda_i|^2}, 
\end{equation}
where $\lambda_i$ are the eigenvalues and $||M||_F=\sqrt{\sum_{i,j=1}^{n}|m_{ij}|^2}$, the Frobenius norm of the matrix $M$ whose elements are $m_{ij}$~\cite{trefethen}. The Henrici departure from normality of the matrix $\hat{A}$ is thus, 
\begin{equation}
    d_f\left(\hat{A}\right)=\frac{\sqrt{\bra{\psi_i}\hat{O}^2\ket{\psi_i}-\bra{\psi_i}\hat{O}\ket{\psi_i}^2}}{|\braket{\psi_f|\psi_i}|^2}=\frac{\Delta_i\hat{O}}{|\braket{\psi_f|\psi_i}|^2}, 
\end{equation}
where $\Delta_i\hat{O}$ is the uncertainty of the observable $\hat{O}$ in the initial state. The details on the derivation of this equation can be found in Appendix \ref{appendix:calculations_henrici_departure_from_normality}. The Henrici departure from normality of $\hat{A}$ vanishes and the matrix is normal, only when the expectation value of $\hat{O}^2$ in the initial state is equal to the square of the expectation of $\hat{O}$ in the initial state, namely
\begin{equation}
    \bra{\psi_i}\hat{O}^2\ket{\psi_i}=\bra{\psi_i}\hat{O}\ket{\psi_i}^2\, .
\end{equation}
However the Cauchy-Schwarz inequality implies that, 
\begin{equation}
    \bra{\psi_i}\hat{O}^2\ket{\psi_i}\geq\bra{\psi_i}\hat{O}\ket{\psi_i}^2,
\end{equation}
where the equality holds true only when $\ket{\psi_i}$ is an eigenvector of $\hat{O}$, $\hat{O}\ket{\psi_i}=\lambda\ket{\psi_i}$. In that case, the operator is normal and the weak value is simply the eigenvalue $\lambda$.

Obviously, defining the operator $\hat{A}$ in terms of the final or the initial state is an arbitrary choice. Hence the weak value can also be defined as the expectation value in the initial state, 
\begin{equation}
 O_w^\prime = \bra{\psi_i}\hat{A}'\ket{\psi_i}\, ,
\end{equation}
 of the operator $\hat{A}'$ given by
 \begin{equation}
 \label{eq:equation_A_prime}
     \hat{A}'=\frac{\hat{\Pi}_f\hat{O}}{\text{Tr}\left[\hat{\Pi}_f\hat{\Pi}_i\right]}.
 \end{equation}

Weak values are called anomalous when the imaginary part is different from $0$ or, otherwise, when their value lies outside the range of the spectrum of the Hermitian operator $\hat{O}$ \cite{pusey2014anomalous, sokolovski2015meaning, ipsen2022anomalous}, namely 
\begin{equation} \Im O_w\neq 0 \hspace{0.5 cm} \text{or} \hspace{0.5 cm} O_w>\text{max}\left(\langle\hat{O}\rangle\right)=\text{max}\left(\text{Tr}\left(\hat{\rho}\hat{O}\right)\right)=\text{max}\left(\lambda_i\right) \hspace{0.5 cm} \text{or} \hspace{0.5 cm} O_w<\text{min}\left(\langle\hat{O}\rangle\right)=\text{min}\left(\text{Tr}\left(\hat{\rho}\hat{O}\right)\right)=\text{min}\left(\lambda_i\right)\, , 
\end{equation}
where $\lambda_i$ are the eigenvalues of the operator $\hat{O}$ and $\hat\rho$ is an arbitrary (not necessarily pure) quantum state represented here as a density operator. Amplifying weak values correspond to $\vert O_w\vert > \max \vert \lambda_i\vert$.

If $\hat{A}$ or $\hat{A}'$ are normal, the weak value cannot be anomalous. Moreover, the weak value is one of the eigenvalues of the observable $\hat{O}$. The operator $\hat{A}$, resp. $\hat{A}'$, have both eigenvalue zero with multiplicity $N-1$ and one eigenvalue equal to an expectation of the observable in the initial state $\bra{\psi_i}\hat{O}\ket{\psi_i}$, resp. final state $\bra{\psi_f}\hat{O}\ket{\psi_f}$. Another interesting case arises when the expectation value is equal to $0$. In such a setting, the weak value is not necessarily equal to zero, but the non-normal matrix has all eigenvalues equal to zero and it is degenerate, hence it is a nilpotent matrix \footnote{The only matrix with a zero spectrum that is not nilpotent is the zero matrix, which is not degenerate, since it has full geometric multiplicity. Our case does not fall in the latter, our matrix being non-normal.}. 

We have thus shown that weak values can be different from an eigenvalue only if both operators $\hat{A}$ and $\hat{A}'$ are non-normal. Since the anomalous properties are fundamental in all applications of weak values, our setting becomes interesting when the involved operators are non-normal.

Beyond this paper specific focus on weak values, we would like to stress that our results also establish a truly general connection between quantum fluctuations and non-normality. Indeed, assuming identical initial and final pure states $\hat\Pi_f=\hat\Pi_i$, the weak value \eqref{eq:weak_value_in_terms_of_trace} is simply the expectation value $\langle \hat{O} \rangle_i$ of the observable in the considered quantum state $\hat\Pi_i$. We can as well express it as the average of the non-normal operator $\hat{A}=\hat{O}\hat\Pi_i$ in the initial state, given simply by the trace $\Tr (\hat{O}\hat\Pi_i)$. Then, Henrici's departure from non-normality becomes exactly equal to the uncertainty of the observable in the quantum state $d_f(\hat{O}\hat\Pi_i)=\Delta_i \hat{O}$. It appears thus that the non-normality of the operator $\hat{O}\hat\Pi_i$ is a measure of the quantum fluctuations around the observable expectation value evaluated through the trace of the same operator $\hat{O}\hat\Pi_i$. When the state $\hat\Pi_i$ is an eigenstate of the observable $\hat{O}$, the operator is normal and the quantum uncertainty is zero. Indeed, measurements of a quantum system in an eigenstate of the probed observable yield the associated eigenvalue with probability 1. This general link between quantum uncertainties and non-normality, and their relationship with Henrici's departure from normality in particular, does not seem to have been recognized in practice, to the best of our knowledge.
\section{Correlation between non-normality and amplifying weak values}\label{sec:numerical_results}
Non-normality, as we have shown in the previous sections, is necessary to obtain weak values different from an eigenvalue of the observable, and thus also to have amplifying and complex weak values. The goal of this section is to study the relation between the level of amplification of the modulus of the weak value, and the non-normality of the matrices $\hat{A}$ and $\hat{A}'$, defined in Eq.~(\ref{eq:definition_matrix_A}) and Eq.~(\ref{eq:equation_A_prime}. We show that there is a direct dependence between those quantities. The larger the amplification, the larger the level of non-normality of matrices $\hat{A}$ and $\hat{A}'$. To measure the non-normality of a matrix, we use the Henrici departure from normality, $d_f$, defined in Eq.~(\ref{eq:Henrici_departure_from_normality}). For the sake of clarity, we consider the case of two-level systems. A general qubit state $\ket{\psi_a}$ can be expressed as
\begin{eqnarray}
\label{eq:general_qubit_state}
\ket{\psi_a}=
\begin{pmatrix}
\cos{\theta_a}\\
e^{i\xi_a}\sin{\theta_a}
\end{pmatrix},
\end{eqnarray}
where $0\leq\theta_a\leq\frac{\pi}{2}$ and $0\leq\xi_a\leq 2\pi$.

The pre- and post-selected states, $\ket{\psi_i}$ and $\ket{\psi_f}$, are described similarly to Eq.~(\ref{eq:general_qubit_state}). Each state, pre- and post-selected, depends on two free parameters and the measured operator depends on four free parameters. Consequently, the full description of the process depends on eight free parameters. In this section, we will vary the pre- and post-selected states and fix the observable, while in the next section, we will do the opposite. We restrict the parametric freedom of the states by imposing the absence of phase of the initial and final states, $\xi_i=0$ and $\xi_f=0$. We refer the interested reader to Appendix~\ref{appendix:some_analytical_formulas} for the analysis of the general case.

To elucidate the correlation between amplifying weak values and non-normality, the modulus of the weak value is qualitatively and quantitatively compared to the Henrici departure from normality. The chosen observables to study are the Pauli matrices and a linear combination of them, $\hat{O}=\frac{1}{\sqrt{3}}\left(\hat{\sigma}_x+\hat{\sigma}_y+\hat{\sigma}_z\right)$, where the Pauli matrices have been defined in Appendix~\ref{appendix:Pauli_matrices}. Our choice has been motivated by the important role the latter play in quantum physics by describing the spin \cite{lemke2009spin, eggert1994susceptibility, marklund2007dynamics}.
\begin{figure}
    \centering
    \includegraphics[width=16cm]{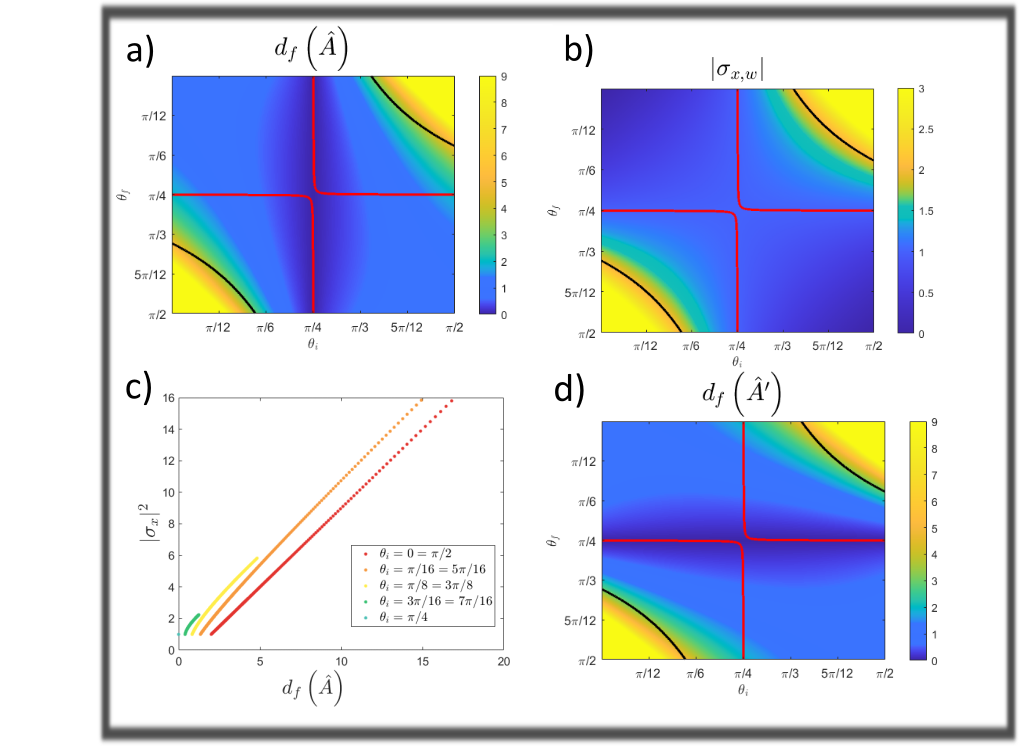}
    \caption{a) Levels set of the modulus of the weak value of $\hat{\sigma}_x$ in terms of the polar angles of the pre- and post-selected states $(\theta_i, \theta_f)$, imposing $\xi_i=0$, $\xi_f=0$. b) Levels set of the Henrici departure from normality of the non-normal matrix $\hat{A}$ associated to the weak value $\sigma_{x,w}$. a,b,d) The red curve corresponds to the border of the area in which the modulus of the weak value is larger than $1.0$. The black curve corresponds to the boundary of the area in which the modulus of the weak value is twice the maximum possible expectation weak value. c) Square modulus of the weak value as a function of the Henrici departure from normality of the operator $\hat{A}$ for anomalous weak values, $|\sigma_{x,w}|>1$, obtained by varying $\theta_f$ between $0$ and $\frac{\pi}{2}$, while $\theta_i$ is fixed (colored dots in the legend). d) Levels set of the Henrici departure from normality of the non-normal operator $\hat{A}'$ associated with the weak value $\sigma_{x,w}$. \color{black}
    \label{fig:figure_sigma_x}}
\end{figure}
\begin{figure}
    \centering
    \includegraphics[width=16cm]{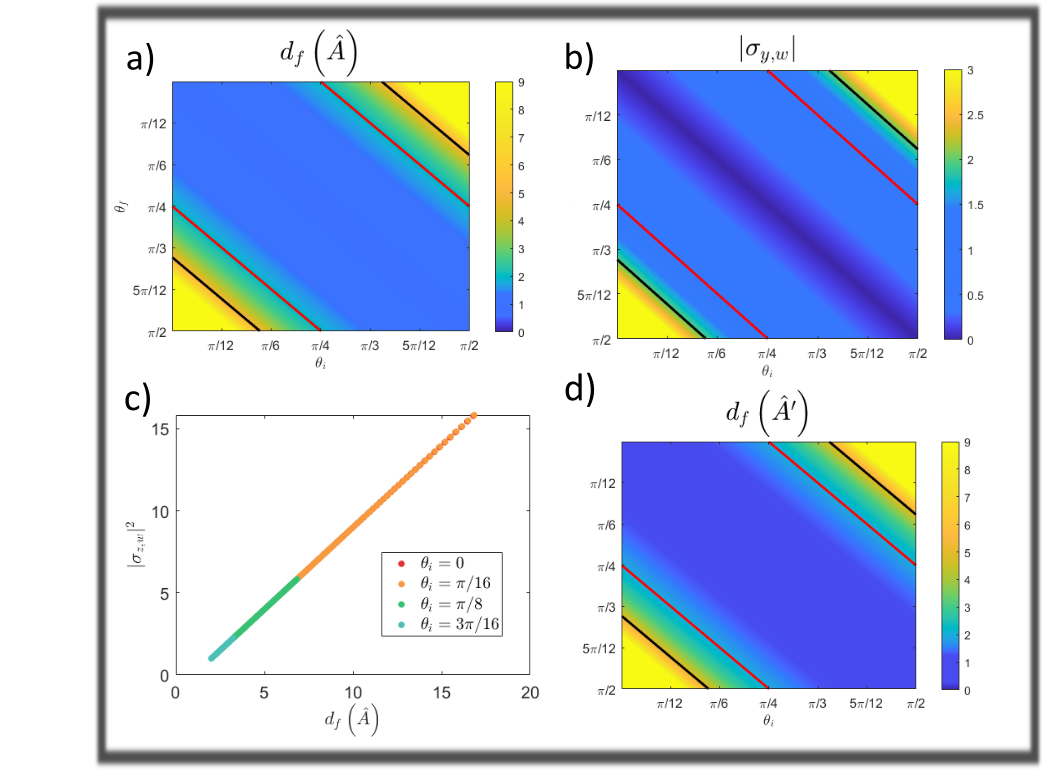}
    \caption{a) Levels set of the modulus of the weak value of $\hat{\sigma}_y$ in terms of the polar angles of the pre- and post-selected states $(\theta_i, \theta_f)$, imposing $\xi_i=0$, $\xi_f=0$. b) Levels set of the Henrici departure from normality of the non-normal matrix associated with the weak value, $\hat{A}$. a,b,d) The red curve corresponds to the boundary of the area in which weak value amplification occurs. The black curve corresponds to the frontier of the area in which the modulus of the weak value is twice the maximum possible expectation value. c) Square modulus of the amplified weak value as function of the Henrici departure from normality of the operator $\hat{A}$, for various $0<\theta_f<\frac{\pi}{2}$. d)  Levels set of the Henrici departure from normality of the non-normal operator $\hat{A}'$ associated with the weak value $\sigma_{y,w}$.}
    \label{fig:figure_sigma_y}
\end{figure}

Let us consider the Pauli matrix $\hat\sigma_x$. Then by using the previous definitions we get 
\begin{equation}
\label{eq:dfsigmax1}
d_f\left(\hat{A}_x\right)=\frac{|\cos(2\theta_i)|}{|\braket{\psi_f|\psi_i}|^2}\, ,
\end{equation}
where $\hat{A}_x$ denotes the operator $\hat{A}$ built from $\hat\sigma_x$. The Henrici departure from normality of the symmetrical operator $d_f\left(\hat{A}'_x\right)$ is
\begin{equation}
\label{eq:dfsigmax2}
d_f\left(\hat{A}'_x\right)=\frac{\Big|\cos(2\theta_f)\Big|}{|\braket{\psi_f|\psi_i}|^2}\, ,
\end{equation}
where $\hat{A}'_x$ denotes the operator $\hat{A}'$ built from $\hat\sigma_x$, and
\begin{equation}
|\sigma_{x,w}|^2=\frac{\sin^2(\theta_f+\theta_i)}{|\braket{\psi_f|\psi_i}|^2}\, ,
\end{equation}
with $|\braket{\psi_f|\psi_i}|^2 = \cos^2(\theta_f-\theta_i)$. The level curves of the latter are reported in Fig.~\ref{fig:figure_sigma_x} as a function of the angles $\theta_i$ and $\theta_f$. The red line define the $1$-level, where the modulus of the weak value equals 1, while the black line denotes the $2$-level. In the region beyond the $1$-level, weak value amplification occurs (as the eigenvalues of Pauli matrices are $\pm 1$). We can observe a very good agreement among the results shown in panels a), b), and d). Since we are considering pre- and post-selected states with real coefficients, the $1$-level curve also define the region beyond which weak values of $\hat\sigma_x$ are anomalous. In other words, all real, anomalous weak values of Pauli matrices provide amplification.

To strengthen this claim we can express $|\sigma_{x,w}|^2$ in terms of $d_f\left(\hat{A}_x\right)$ by eliminating, e.g., the variable $\theta_f$ and considering thus $\theta_i$ as a free parameter. We can thus obtain
\begin{equation}
|\hat{\sigma}_{x,w}|^2=\frac{1}{1+\tan^2 \theta_i}\frac{1}{1+\tan^2 \theta_f}\frac{(\tan \theta_f+\tan\theta_i)^2}{|\cos(2\theta_i)|}d_f\, ,
\end{equation}
where $\tan \theta_f$ can be expressed as a function of $d_f$ by using~\eqref{eq:dfsigmax1}. The explicit formula can be found in Appendix~\ref{appendix:some_analytical_formulas}. This relation is shown in panel c) of Fig.~\ref{fig:figure_sigma_x} for several values of the parameter $\theta_i$ and by restricting $d_f$ to the range corresponding to anomalous weak values, i.e., $|\hat{\sigma}_{x,w}|^2>1$.

In conclusion in Fig.~\ref{fig:figure_sigma_x}, we have considered the case for the observable $\hat{\sigma}_x$ and observed that the Henrici departure from normality of both $\hat{A}$ and $\hat{A}'$ (panels a and d) and the pattern of the modulus of the weak value (panel b) exhibit a good correlation in the region of anomalous weak values, especially in the region where $|O_w|>2$, namely associated to yellow values. Near the boundary between anomalous and non-anomalous weak values (red curves), some differences are appreciable: in particular, the value of the minimum of the Henrici departure from normality of both $\hat{A}$ and $\hat{A}'$ are smaller than the modulus of the weak value (the former are associated with darker blue regions than the latter). On the other hand, considering the region associated to values that are not anomalous, i.e., the region bounded by the red curve and containing the point $(\pi/4,\pi,4)$, the patterns are completely different, we can indeed appreciate the presence of minima in the top-left and bottom-right corners for the weak value (b) and in vertical and horizontal center lines in the case of the Henrici departure from normality of $\hat{A}$ and $\hat{A}'$ (a and d). 
In Fig.~\ref{fig:figure_sigma_x}c), we observed that the squared modulus of the weak value increases as a function of the Henrici departure from normality. The dependence appears linear once the weak value reaches large values (red and orange curves). The point of interception with the axis $\left(|\sigma_{x,w}|^2=0\right)$ depends on the initial angle, $\theta_i$. However, when the weak value does not reach large values (green and yellow curves), the dependence is closer to quadratic. The analytical formulas can be found in appendix \ref{appendix:some_analytical_formulas}. The behaviour of these functions is thus very rich.

\begin{figure}
    \centering
    \includegraphics[width=16cm]{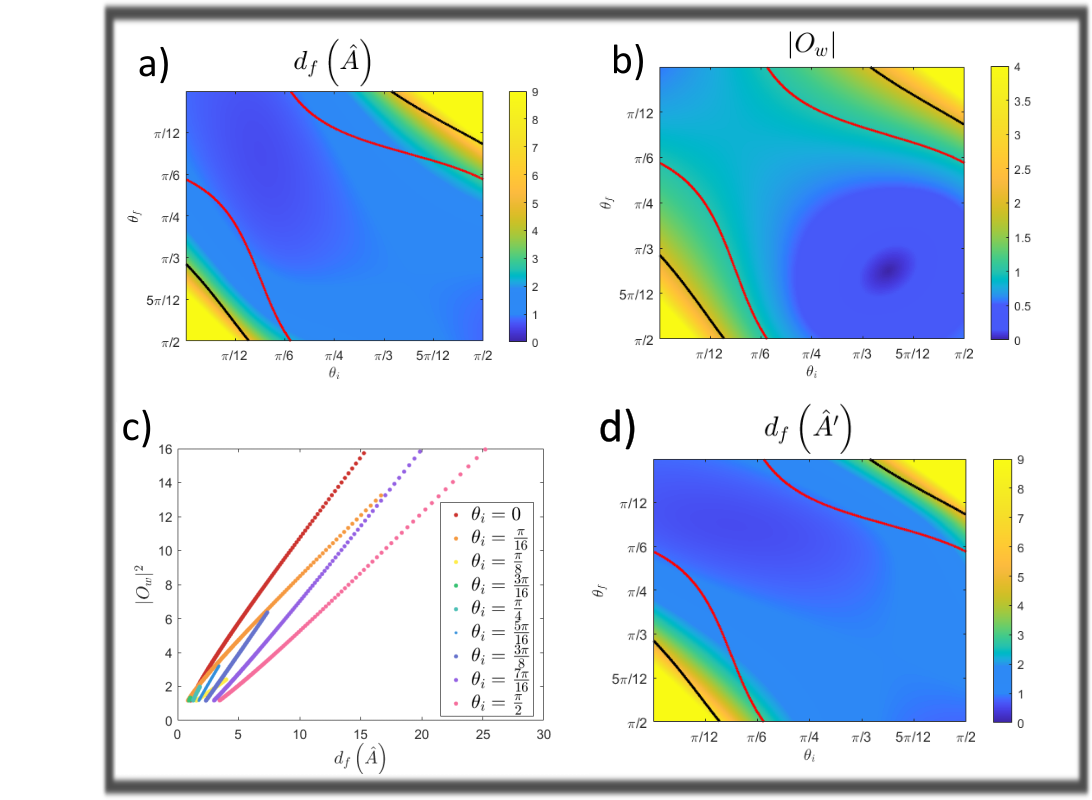}
    \caption{a) Levels set of the modulus of the weak value of the operator $\hat{O}=\frac{1}{\sqrt{3}}\left(\hat{\sigma}_y+\hat{\sigma}_x+\hat{\sigma}_z\right)$ in terms of the polar angles of the pre- and post-selected states $(\theta_i, \theta_f)$, imposing $\xi_i=0$, $\xi_f=0$. b) Levels set of the Henrici departure from normality of the non-normal matrix $\hat{A}$ associated with the weak value in terms of the polar angles of the pre- and post-selected states $(\theta_i, \theta_f)$. a,b,d) The red curve corresponds to the boundary of the area in which the weak value amplification occurs. The black curve corresponds to the frontier of the area in which the modulus of the weak value is twice the maximum possible expectation value. c) Square modulus of the weak value as a function of the Henrici departure from normality for amplified anomalous weak values, $|O_{w}|>1$, with $0<\theta_f<\frac{\pi}{2}$. d) Levels set of the Henrici departure from normality of the non-normal operator $\hat{A}'$ associated with the weak value $O_{w}$.} \label{fig:figure_sigma_x_plus_sigma_y_plus_sigma_z}
\end{figure}

A similar analysis can be performed by using as observables the remaining Pauli matrices and a combination of them. In the following Figs.~\ref{fig:figure_sigma_y},~\ref{fig:figure_sigma_z},~\ref{fig:figure_sigma_x_plus_sigma_y_plus_sigma_z}, we report the dependence of the modulus of the weak value of $\hat{\sigma}_y$, $\hat{\sigma}_z$ and $\hat{O}=\frac{1}{\sqrt{3}}\left(\hat{\sigma}_x+\hat{\sigma}_y+\hat{\sigma}_z\right)$ respectively, as a function of the parameters $\theta_i$ and $\theta_f$ by using adapted color maps to emphasize the level curves. In the same figures, we also show the Henrici departure from normality of the operators $\hat{A}$ and $\hat{A}'$ as a function of the same parameters. As in the case of the observable $\hat{\sigma}_x$, we can observe (see panels a), b) and d) in all the figures) a very good agreement between the square of the modulus of the weak value and the Henrici departures from normality. We have also shown the region in which the modulus of the weak value is larger than the absolute value of the largest eigenvalue of the observable (red curve) and the one for which the modulus is at least twice the maximum eigenvalue (black curve). Outside the regions bounded by the latter curves, i.e., the yellow regions, the weak value is amplifying. Inside the regions bounded by the red curve, i.e., the blue ones, the modulus of the weak value lies in the range of eigenvalues of the studied operator (but the weak value can still be anomalous, i.e. if it is a complex number).
\begin{figure}
    \centering
    \includegraphics[width=16cm]{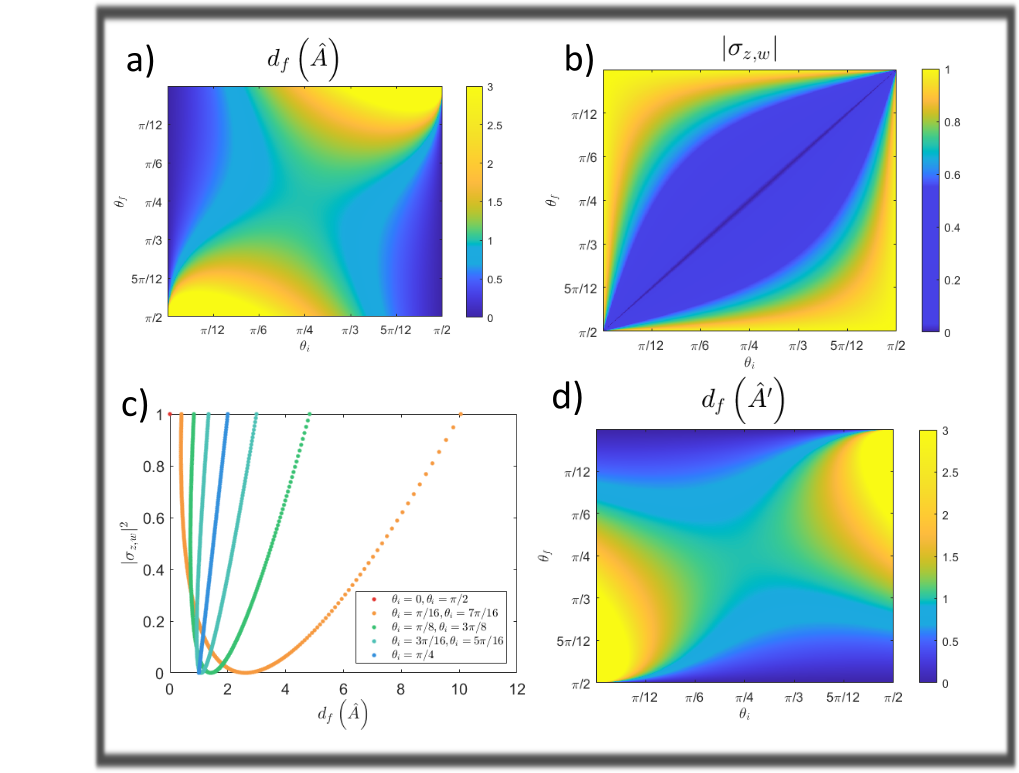}
    \caption{a) Levels set of the modulus of the weak value of $\hat{\sigma}_z$ in terms of the polar angles of the pre- and post-selected states $(\theta_i, \theta_f)$, imposing $\xi_i=0$, $\xi_f=0$. b) Levels set of the Henrici departure from normality of the non-normal matrix $\hat{A}$ associated with the weak value. c) Square modulus of the weak value as a function of the Henrici departure from normality, with $0<\theta_f<\frac{\pi}{2}$. d) Levels set of the Henrici departure from normality of the non-normal operator $\hat{A}'$ associated with the weak value $\sigma_{z,w}$.}
    \label{fig:figure_sigma_z}
\end{figure}

In Fig.~\ref{fig:figure_sigma_y}, the chosen observable is $\hat{\sigma}_y$ and we report again the weak value and the Henrici departure from normality of $\hat{A}$ and $\hat{A}'$ as a function of the angles by using a color code match. In this case too, we have a perfect agreement of the behavior of the three quantities in the parameter region associated to anomalous weak values. In the complementary region, a central minimum appears in the case of the weak value that is absent in the plots of the Henrici departures from normality. The results presented in Fig.~\ref{fig:figure_sigma_y}c), show that the square of the modulus of the weak value depends linearly on the Henrici departure from normality of $\hat{A}$, irrespective of the magnitude of the maximum weak value for the specific case. Furthermore, the slope and the point of interception with the axis $\left(|\sigma_{y,w}|^2=0\right)$ are independent on the value of $\theta_i$. In appendix \ref{appendix:analytical_calculations_considering_null_phases}, one can find the analytical formula showing that the modulus of weak value to the square depends linearly on the Henrici departure from normality, when both phases are null. Note that in Fig.~\ref{fig:figure_sigma_y}b), all non-zero weak values are anomalous because they are purely imaginary numbers (anomalousness is not equivalent to amplification in this case).

In Fig.~\ref{fig:figure_sigma_z}, where the chosen observable is $\hat{\sigma}_z$, the patterns of the three quantities do not exhibit similarities. This difference with respect to the other cases is due to the fact that no amplification occurs for this observable with the chosen pre- and post-selected states. The weak value is real and never anomalous. In Fig.~\ref{fig:figure_sigma_z}c), we have plotted the dependence of the modulus of the weak value on the Henrici departure from normality of $\hat{A}$ for many values of the angles. As one can see, a large weak value does not imply a large Henrici departure from normality, indeed there is a parabola-like behavior. The analytical formulas can also be found in appendix \ref{appendix:analytical_calculations_considering_null_phases}.

In Fig.~\ref{fig:figure_sigma_x_plus_sigma_y_plus_sigma_z}, we show the results for the observable $\hat{O}=\frac{1}{\sqrt{3}}(\hat{\sigma}_x+\hat{\sigma}_y+\hat{\sigma}_z)$. The pattern of the modulus of the weak value and the Henrici departures from normality show a good correlation in the region of weak value amplification, especially where $|O_w|>2$ (black curve). Some differences can be appreciated near the boundary (red curves) separating the regions where amplification occurs or not, as the weak value and the departure from normality of $\hat{A}$ and $\hat{A}'$ exhibit different patterns near that area. In the complementary region, i.e., inside the region bounded by the red curves and containing the point $(\pi/4,\pi/4)$, large differences can be appreciated: the level sets for the weak value show a ring-like shape absent in the plots of the Henrici departures from normality. In Fig.~\ref{fig:figure_sigma_x_plus_sigma_y_plus_sigma_z}c), there is a linear-like dependence of the square of the modulus of the weak value on the Henrici departure from normality. However, in this case, both the point of interception of the axis $\left(|O_w|^2=0\right)$ and the slope depend on $\theta_i$. When the modulus of the weak value does not reach large values (green and blue lines), the dependence is quadratic and not linear. For this observable and the chosen pre- and post-selected states, the weak value is a complex number whenever $\theta_i \neq \theta_f$, and thus anomalous everywhere but on the descending diagonal of Fig.~\ref{fig:figure_sigma_x_plus_sigma_y_plus_sigma_z}b).

\begin{figure}
    \centering
    \includegraphics[width=16cm]{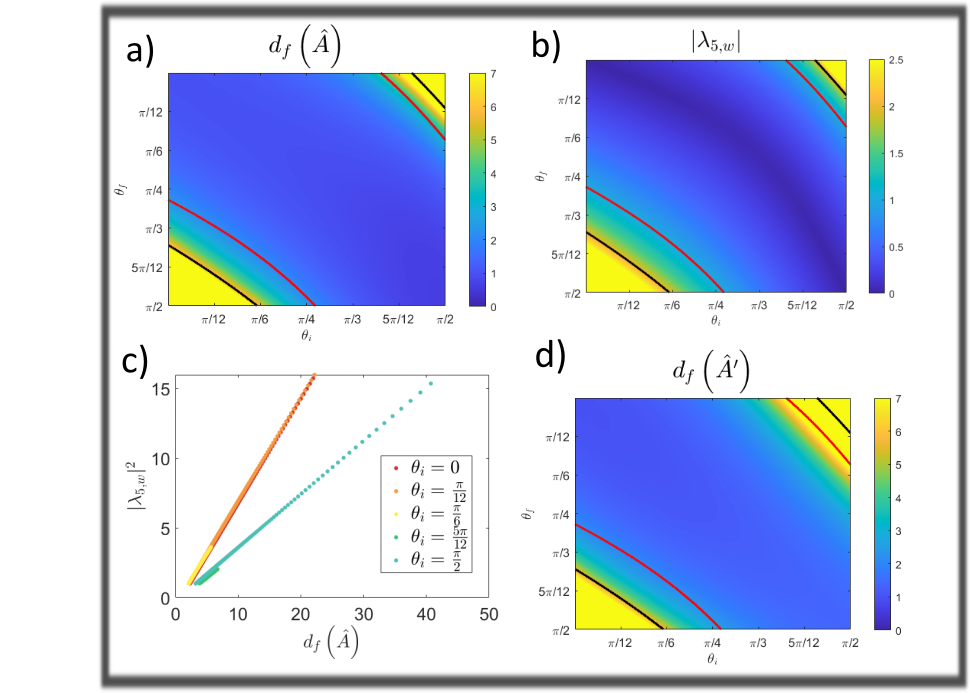}
    \caption{a) Levels set of the modulus of the weak value of the operator $\hat{O}=\hat{\lambda}_5$, $\chi_{1,i}=\frac{\pi}{7}$, $\chi_{2,i}=\frac{\pi}{21}$, $\alpha_{i}=\frac{\pi}{8}$, $\chi_{1,f}=\frac{\pi}{4}$, $\chi_{2,f}=0$, $\alpha_{f}=\frac{\pi}{3}$ in terms of the polar angles of the pre- and post-selected states ($\theta_i$ and $\theta_f$), imposing $\xi_i=0$, $\xi_f=0$. b) Levels set of the Henrici departure from normality of the non-normal matrix $\hat{A}$ associated with the weak value. a,b,d) The red curve corresponds to the boundary of the area in which the weak value is in the amplification range. The black curve corresponds to the frontier of the area in which the modulus of the weak value is twice the maximum possible expectation value. c) Square modulus of the weak value as function of the Henrici departure from normality for amplifying weak values, $|O_{w}|>1$, having $0<\theta_f<\frac{\pi}{2}$. d) Levels set of the Henrici departure from normality of the non-normal matrix $\hat{A}'$ associated with the weak value.}
\label{fig:figure_three_level}
\end{figure}

For the sake of completeness, we have also computed the weak value and the Henrici departures from normality for a three-level system, whose general state is 
\begin{equation}
    \ket{\psi_a}=\left(\cos{\theta_a}, e^{i\chi_{1,a}}\cos{\alpha_a}\sin{\theta_a}, e^{i\chi_{2,a}}\sin{\alpha_a}\sin{\theta_a}\right)^\top\, .
\end{equation}
The chosen three-level operator is the Gell-Mann matrix $\hat{O}=\hat{\lambda}_5$ (see Appendix~\ref{appendix:gell_mann}). The Gell-Mann matrices are the traceless generators of the Lie group $\textrm{SU}(3)$ that generalize the Pauli matrices, the traceless generators of $\textrm{SU}(2)$. In Fig.~\ref{fig:figure_three_level}, the weak value and the Henrici departure from normality of $\hat{A}$ and $\hat{A}'$ have been depicted by using a color map scheme. As one can see, the plots match pretty well, with the exception of the angles located in the top-right corner, for which the Henrici departures from normality, specially the one of $\hat{A}'$, assume larger values than the squared modulus of the weak value, in particular in the strip contained between the red and black curves. The square of the modulus of the weak value depends linearly on the Henrici departure from normality. The slope and the point of interception $\left(|\lambda_{5,w}|^2=0\right)$ depend on the pre-selected polar angle. 

In conclusion, we have shown that the amplification degree of the weak value depends on the non-normality of the matrices $\hat{A}$ and $\hat{A}'$ when weak value amplification occurs (so that the weak value is necessarily anomalous, even if a real number). When one feature increases, the other does too. However, this does not happen when amplified weak values are not present. In general, the behaviour is complex, and there are cases in which some regions of discordance are present, as shown in appendix \ref{appendix:some_analytical_formulas}.

\section{A reformulation of the problem with a varying observable}\label{new_expression}
In the previous section, we examined the relationship between the Henrici departures from normality and the weak value by varying the pre- and post-selected states. However, in some cases, we may have the freedom to select the observable for a given experiment. Here, we investigate how varying the observable impacts non-normality and the modulus of weak values, for given pre- and post-selected states. To narrow our focus, we restrict our analysis to two-level systems. Specifically, we consider an observable that depends on two parameters,
\begin{equation}
\label{eq:varying_operator}
    \hat{O}=\sin{\theta}\cos{\phi}\ \hat{\sigma}_x+\sin{\theta}\sin{\phi}\ \hat{\sigma}_y+\cos{\theta}\ \hat{\sigma}_z\, , 
\end{equation}
where $0\leq \theta \leq \frac{\pi}{2}$, $0\leq \phi \leq 2\pi$, and $\hat{\sigma}_i$ are the Pauli matrices (Appendix~\ref{appendix:Pauli_matrices}). In the present analysis, we investigate the relationship between the weak value and the Henrici departure from normality for the two matrices $\hat{A}$ and $\hat{A}'$, defined in Eq.~\eqref{eq:definition_matrix_A} and Eq.~\eqref{eq:equation_A_prime}, respectively. Because the denominators of the latter matrices do not vary by modifying the operator, we decided to compare, in this section, the normalized Henrici departures from normality -- $d_{f,n}\left(\hat{A}\right)$ and $d_{f,n}\left(\hat{A}'\right)$ -- with the modulus of the numerator of the weak value, i.e., $|\bra{\psi_f}\hat{O}\ket{\psi_i}|$. \\

\begin{figure}
    \centering
    \includegraphics[width=8cm]{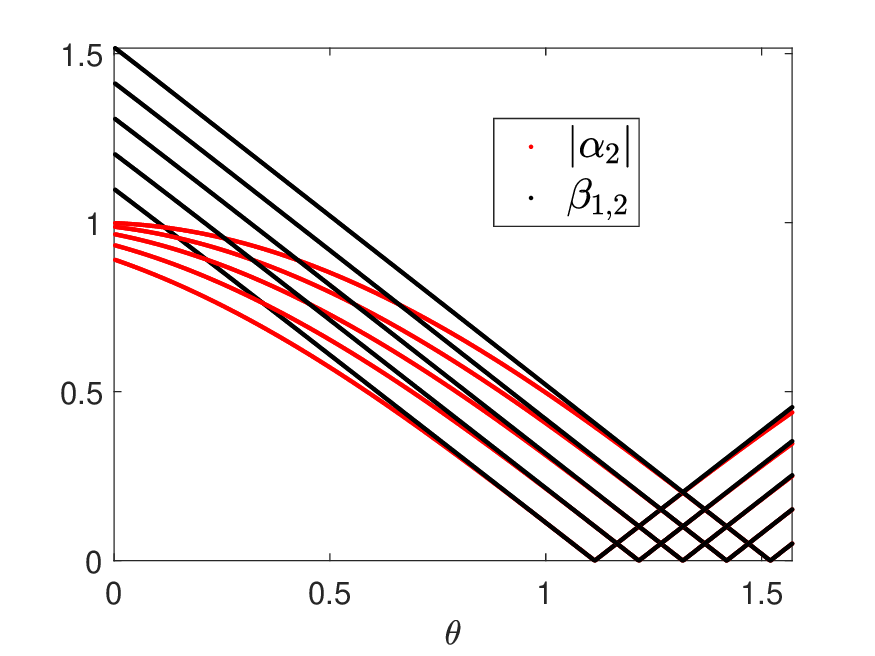}
    \caption{
    For any given polar initial angle, a completely degenerate nilpotent point exists. The largest eigenvalue in absolute value of $\hat{A}$, $\alpha_2$, has been plotted in terms of $\theta$ and the angle between the eigenvectors ($\vert \beta_1\rangle$ and $\vert \beta_2\rangle$) of $\hat{A}$, $\beta_{1,2}=\arccos{\left\vert \langle\beta_1 \vert \beta_2\rangle\right\vert}$, the Fubini-Study distance, in terms of $\theta$. The chosen parameters are $\phi=\frac{\pi}{12}$, $\theta_f=0$, $\xi_f=0$, $\xi_i=0$. The polar angle of the initial state varies between $\frac{\pi}{2}$ and $\frac{5\pi}{12}$.\label{fig:eigenvalue_and_angle_eigenvector}}
\end{figure}

In this setting, for very large weak values, we find that the amplification grows with the non-normality, analogously to the behavior shown in the yellow regions of the Figures in the previous section (more details are provided in Appendix~\ref{appendix:sec4_new}, Fig. \ref{fig:figure_varying_operator_first_case} in particular). However, a much richer behavior occurs in the low amplification range (similar to the regions between the red and black lines in the Figures of the previous section), where the normalized Henrici departures from normality might increase while the modulus of the weak value decreases. The latter results, together with further details on the analysis, can be found in Appendix \ref{appendix:sec4_new}. Specifically, we show there how the modulus of the weak value varies with the non-normality of $\hat{A}$ and $\hat{A}'$, as expressed by their normalized Henrici departure from normality.
As already discussed, both operators $\hat{A}$ and $\hat{A}'$ have an eigenvalue equal to zero and the other one equal to $\frac{\bra{\psi_i}\hat{O}\ket{\psi_i}}{\braket{\psi_f|\psi_i}}$ and $\frac{\bra{\psi_f}\hat{O}\ket{\psi_f}}{\braket{\psi_f|\psi_i}}$, respectively. Remarkably, the parameter value ($\theta$) for which one of these non-normal operators (i.e., the $2\times 2$ matrix in a determined basis) is nilpotent is also such that the associated expectation value of the observable $\hat{O}$ is equal to $0$. This is an interesting observation because it is well known that nilpotent matrices and operators affect the system dynamics \cite{nilp1,nilp2}. Furthermore, the $0$ eigenvalue is now degenerate, as one can observe from Fig.~\ref{fig:eigenvalue_and_angle_eigenvector}, where the largest eigenvalue in absolute value and the angle between eigenvectors, calculated from the Fubini-Study distance, have been plotted for different initial states. Let us observe that the algebraic multiplicity of the $0$ eigenvalue is $2$, but its geometric multiplicity is $1$ (there is a single eigenvector, hence the eigenspace is spanned by a single nonzero vector), meaning that the matrix is degenerate, and hence non-normal~\footnote{One may recall that the only $2\times 2$ matrix with $0$ eigenvalue of algebraic multiplicity equal to $2$ that is not degenerate is the null matrix.}.  As expected, for any value of the initial state, when the largest eigenvalue in absolute value is $0$ (nilpotent matrix), the angle between the eigenvectors is null, due to the fact that the matrix is degenerate.

\begin{figure}
    \centering
    \includegraphics[width=8cm]{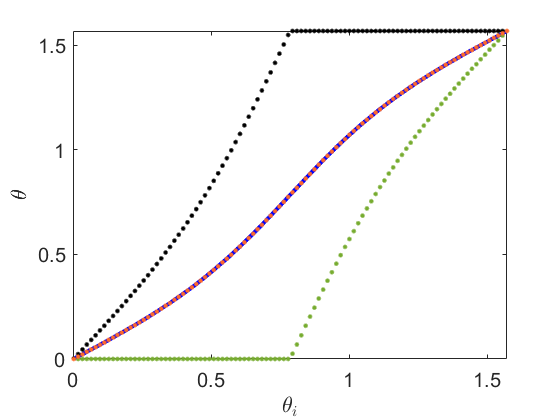}
    \caption{Value of $\theta$ from the observable $\hat{O}$ corresponding to the minimum of the largest eigenvalue in absolute value of $\hat{A}$, $\alpha_2$ (green) and $\hat{A}'$, $\alpha'_2$ (black), and to the maximum modulus of the weak value (blue) as a function of $\theta_i$, the polar angle of the pre-selected state, for anomalous weak values in the amplification range. The average of $\theta$ corresponding to the minima of the maximum of the absolute value of eigenvalue of both matrices (orange) has also been included. We can observe that the blue and the orange curves coincide, meaning that the maximum modulus of the numerator of the weak value is attained at the average of the points in which the normalized Henrici indexes reach their maxima for both matrices.
     The chosen parameters are $\phi=\frac{\pi}{4}$, $\theta_f=0$, $\xi_f=0$, $\xi_i=0$. \label{fig:angle_nilpotent_angle}}
\end{figure}

Furthermore, the point (value of $\theta$) of the maximum modulus of the numerator of the weak value in the amplification regime, i.e., when the modulus of the weak value is larger than the maximum absolute value of the eigenvalues of the operator), is completely determined by the two points where the normalized Henrici departure from normality of the matrices $\hat{A}$ and $\hat{A}'$ reach their maximum values. Indeed, the maximum modulus of the numerator of the weak value is always attained at the average of the two points where the normalized Henrici departure from normality reaches its maximum for the two matrices (it is also the point in which the largest eigenvalue in absolute value reaches its minimum), as shown in Fig.~\ref{fig:angle_nilpotent_angle}, where one can observe that the dotted blue curve (value of $\theta$ at the maximum modulus of the numerator of the weak value) coincides with the plain orange one (average of values of $\theta$ where the normalized Henrici indexes of the two matrices reach their maxima). More details of these results can be found in Appendix~\ref{appendix:maximum_weak_value_maximum_df}.

\begin{figure}
    \centering
    \includegraphics[width=8cm]{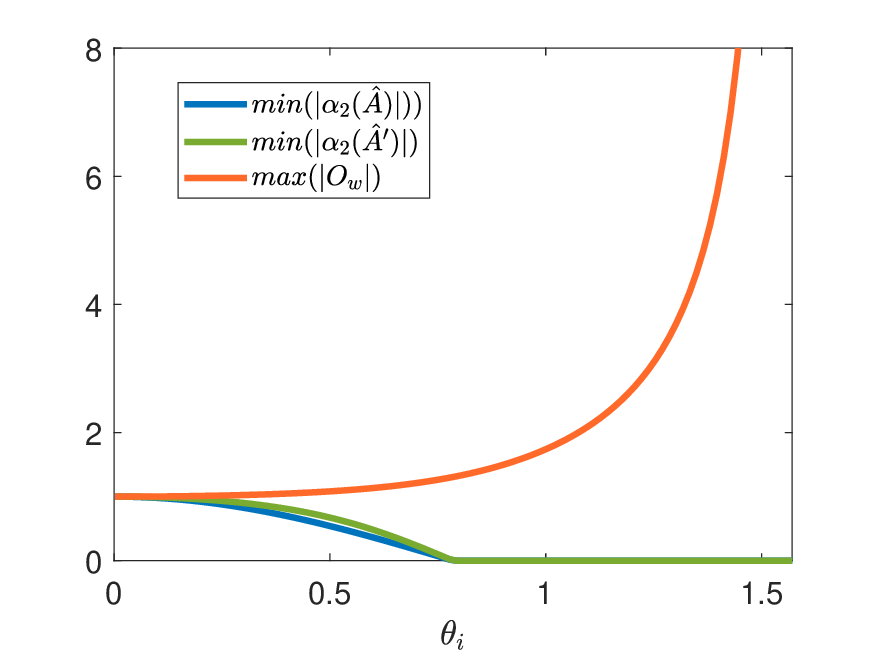}
    \caption{Largest value of the modulus of the weak value (orange), smallest value of the largest eigenvalue in absolute value of $\hat{A}$, $\alpha_2$ (blue) and $\hat{A}'$, $\alpha'_2$(green), as a function of the initial state ($\theta_i$). The extreme values are computed over the range of all allowed values for the angle $\theta$ associated with the varying observable $\hat{O}$. We can observe that the maximum of the modulus of the weak value starts to grow when the degeneracy points are the same and, moreover, that it tends to infinity when the points of zero eigenvalues occur at the same angle $\theta$. The chosen parameters are $\phi=\frac{\pi}{4}$, $\theta_f=0$, $\xi_f=0$, $\xi_i=0$. \label{fig:weak_value_and_henrici}}
\end{figure}

As the angle $\theta_i$ increases from $0$ to $\pi/4$, the range of the parameter $\theta$ in which weak value amplification occurs becomes more pronounced. Starting from $\theta_i=\frac{\pi}{4}$, the degeneracy point, i.e., where both eigenvalues of $\hat{A}$ and $\hat{A}'$ are equal to zero, is located inside the amplification region, as shown in Fig.~\ref{fig:weak_value_and_henrici}. Until this point, the maximum modulus of the weak value is smaller than twice the largest eigenvalue of $\hat{O}$, meaning that we are in the weak amplification regime (between red and black lines in the figures of the previous section). However, as $\theta_i$ increases beyond $\frac{\pi}{4}$, the maximum modulus of the weak value begins to rise much more steeply. In that region, the closer the degeneracy points of the eigenvalues of the operators $\hat{A}$ and $\hat{A}'$ approach to each other, the larger the maximum of the modulus of the weak value. Another relevant fact is that the maximum of the modulus of the weak value tends to infinity when the degeneracy points coincide (i.e., located at the same value of $\theta$).  

For a given two-level quantum state on the Bloch sphere, there are infinitely many Pauli operators with zero average value in that state. They are all located on the great circle whose normal is given by the state in which the average is evaluated. Thus, for any given combination of pre- and post-selected states, there are always two operators for which both $\hat{A}$ and $\hat{A}’$ are degenerate (the intersection of the two corresponding great circles).  As the pre- and post-selected states approach orthogonality, the two great circles on the Bloch sphere move closer to each other. As a result, for the given one-dimensional parametrization of the operator as a function of $\theta$, the intersections of the $\theta$-varying operator curve with the two great circle become closer to each other. When the pre- and post-selected states become perfectly orthogonal, the two great circles coincide, resulting in a complete circle of operators exhibiting degeneracy with respect to both states simultaneously. Consequently, the theta-varying operator curve intersects the two degenerate circles at the same point, effectively reducing the distance between these two points of intersection to zero. We note in passing that the orthogonality of the pre- and post-selected states is necessary for the weak value to become infinite (the degeneracy of both $\hat{A}$ and $\hat{A}’$ is not sufficient in itself) .

In summary, the closer the degeneracy points of the eigenvalues of $\hat{A}$ and $\hat{A}'$ are to each other, the larger the maximum modulus of the weak value becomes. Ultimately, the maximum modulus of the weak value tends to infinity when the points of zero eigenvalues occur at the same angle $\theta$.

\section{Discussion and conclusions}\label{sec:conclusions}
In this paper, we have shown the connection between non-normality and anomalous weak values, with particular focus on the amplification regime. Unlike expectation values, weak values can be complex or even outside the range of eigenvalues of the operator under scrutiny, i.e., larger than the maximum eigenvalue or smaller than the minimum eigenvalue of the measured operator, $\hat{O}$. In the latter cases, we say that the weak value is anomalous. When such anomalous value has a modulus larger than the largest modulus of the eigenvalues, we reach the amplification regime.

Firstly, we proved that weak values can be expressed as the expectation value of matrix, $\hat{A}$, Eq.\eqref{eq:weak_value_in_terms_of_A} and Eq.\eqref{eq:definition_matrix_A}, which is typically non-normal. Then, we have shown that, indeed, such a matrix must be non-normal to obtain a weak value different from an eigenvalue of the observable, i.e., non-normality is a necessary condition to be in the anomalous regime. There are two arbitrary ways to define such non-normal matrix: either in terms of the initial, i.e., pre-selected, state or of the final, i.e., post-selected, state. Both matrices, called here $\hat{A}$ and $\hat{A}'$, must be non-normal in order to yield an anomalous weak value. The non-normality is linked to the quantum fluctuations of the observable in the pre- or post-selected states. Interestingly, the quantum uncertainty of an arbitrary observable of any system in a given quantum state is exactly equal to Henrici's departure from normality of a specific operator, naturally built from the product of the observable operator with the quantum state projector. This result highlights the importance of non-normality in quantum systems, far beyond the specific context of weak measurements.

We showed for two-level systems that the two matrices $\hat{A}$ and $\hat{A}'$ can become simultaneously nilpotent, i.e., all the eigenvalues are null and both matrices are degenerate. For a given observable, this point is found by searching appropriate pre- and post-selected states. In the case of 2-level systems, the matrices are completely degenerate.

We have derived the analytical expressions for the weak value and the Henrici departure from normality in the relevant case of Pauli matrices and we have numerically compared the modulus of weak values and non-normality when varying the pre- and post-selected states. We have found a good correlation between such quantities once parameters allow for anomalous weak values firmly in the amplification regime. 

Eventually, we have also analyzed the relation between the normalized Henrici departure from normality of both non-normal matrices $\hat{A}$ and $\hat{A}'$ when considering a $1$-parameter family of observables in the weak-value amplification regime. In this case, the point of the maximum modulus of the numerator of the weak value is reached at the arithmetic average of the two points where the Henrici departure from normality of the matrices $\hat{A}$ and $\hat{A}'$ are maximal.

Non-Hermitian Hamiltonians describe the evolution of systems that are not isolated \cite{moiseyev2011non}.
Even though in weak measurements the Hamiltonian of the weak process is Hermitian, post-selection seems to introduce non-normality in the process. It is interesting to observe the analogy with networks, where it is conjectured that non-normality (namely the presence of source and sink nodes) is correlated to the fact that the modeled system is an open one \cite{asllani_leaders}. In weak measurements, it is clear that post-selection plays an important role by transforming an expectation values into weak values that can become anomalous. A reconsideration of weak measurements by viewing them as open rather than closed processes appears necessary.

Moreover, the fact that a non-normal matrix cannot be diagonalized through an orthonormal transformation means that the information encoded in the system cannot be completely disentangled. This is particularly interesting thinking about the present framework, as when post-selection is executed on the system, all the ancilla wave-function shifts are projected on a common state and thus interfere. This interference seems to mix the information of the different shifts in a disentangled manner. To do so, the expectation value of the operator is transformed into a weak value that, as we have seen in this paper, is strongly correlated to the non-normality of the operators $\hat{A}$ and $\hat{A}'$ in the amplification regime.

Lastly, let us recall that, in other fields, it has been observed that non-normal operators enhance the possibility of phase transitions to occur \cite{jtb,entropy}. 
Since phase transitions can be observed in post-selected measurements \cite{zhang2021observation}, we believe that revisiting quantum systems from this new perspective could lead to the appearance of new, unexploited quantum phenomena linked to anomalous weak values. 

\subsubsection*{Authors contributions} 
Initial Idea and Conceptualization: L.B.F.; Methodology: L.B.F., R.M., T.C., Y.C.; Formal Analysis and Investigation: L.B.F., R.M., Y.C., T.C.; Numerical Implementation: L.B.F.; Writing - original draft preparation: L.B.F., R.M; Writing - review and editing: L.B.F., R.M., T.C., Y.C.

\subsubsection*{Acknowledgments} R.M. is supported by a FRIA-FNRS PhD fellowship, Grant FC 33443, funded by the Walloon region. Y.C. is a research associate of the Fund for Scientific Research (F.R.S.-FNRS). This research was supported by the Action de Recherche Concertée WeaM at the University of Namur (19/23-001).

\appendix
\section{Calculation of the expression of the Henrici departure from normality}\label{appendix:calculations_henrici_departure_from_normality}
Let us consider the non-normal operator $\hat{A}$ defined in Eq.\eqref{eq:definition_matrix_A}. The square of this operator is 
\begin{equation}
    \hat{A}^2=\frac{1}{\text{Tr}\left[\hat{\Pi}_f\hat{\Pi}_i\right]^2}\hat{O}\ket{\psi_i}\bra{\psi_i}\hat{O}\ket{\psi_i}\bra{\psi_i}= \frac{\bra{\psi_i}\hat{O}\ket{\psi_i}}{|\braket{\psi_f|\psi_i}|^2}\hat{A},
\end{equation}
which implies that the $N^{\textrm{th}}$ power of the operator $\hat{A}$ is 
\begin{equation}
    \hat{A}^N= \frac{\bra{\psi_i}\hat{O}\ket{\psi_i}^{N-1}}{|\braket{\psi_f|\psi_i}|^{2\left(N-1\right)}}\hat{A}.
\end{equation}
The matrix $\hat{A}$ is almost idempotent, as $ \hat{A}^2=\alpha\hat{A}$. Consequently, we can redefine a matrix $\tilde{\hat{A}}$ that is idempotent, i.e. $\tilde{\hat{A}}^2=\tilde{\hat{A}}$,
\begin{equation}
   \tilde{\hat{A}}=\frac{\hat{O}\hat{\Pi}_i|\braket{\psi_f|\psi_i}|^2}{\bra{\psi_i}\hat{O}\ket{\psi_i}}.
\end{equation}
The eigenvalues of an idempotent matrix are either $0$ or $1$. As the trace of the matrix $\tilde{\hat{A}}$ is $1$, one eigenvalue should be $1$ and the rest $0$. Consequently, one eigenvalue of the matrix $\hat{A}$ is $\frac{\bra{\psi_i}\hat{O}\ket{\psi_i}}{\braket{\psi_f|\psi_i}^2}$ and the rest are 0. Obviously, when the expectation value of the observable in the initial state is $0$, all eigenvalues are $0$.
The Frobenious norm of the matrix $\hat{A}$ is
\begin{equation}
    ||\hat{A}||_F^{2}=\frac{1}{|\braket{\psi_f|\psi_i}|^4}\text{Tr}\left[\left(\hat{O}\hat{\Pi}_i\right)\left(\hat{O}\hat{\Pi}_i\right)^{\dagger}\right]=\frac{1}{|\braket{\psi_f|\psi_i}|^4}\text{Tr}\left[\hat{O}\hat{\Pi}_i\hat{\Pi}_i\hat{O} \right]=\frac{1}{|\braket{\psi_f|\psi_i}|^4}\text{Tr}\left[\hat{O}\hat{\Pi}_i\hat{O} \right]=\frac{\bra{\psi_i}\hat{O}^2\ket{\psi_i}}{|\braket{\psi_f|\psi_i}|^4} \, .
\end{equation}
The Henrici departure from normality of the matrix $\hat{A}$ is
\begin{equation}
d_f\left(\hat{A}\right)=\sqrt{\frac{\bra{\psi_i}\hat{O}^2\ket{\psi_i}}{|\braket{\psi_f|\psi_i}|^4}-\frac{\bra{\psi_i}\hat{O}\ket{\psi_i}^2}{|\braket{\psi_f|\psi_i}|^4}}=\frac{\Delta_i\hat{O}}{|\braket{\psi_f|\psi_i}|^2}\, ,
\label{eq:dfA}
\end{equation}
which is proportional to the uncertainty of the operator $\hat{O}$ in the initial state $\ket{\psi_i}$. 
Following the same procedure, one can obtain the Henrici departure from normality of the operator $\hat{A}'$, 
\begin{equation}
d_f\left(\hat{A}'\right)=\sqrt{\frac{\bra{\psi_f}\hat{O}^2\ket{\psi_f}}{|\braket{\psi_f|\psi_i}|^4}-\frac{\bra{\psi_f}\hat{O}\ket{\psi_f}^2}{|\braket{\psi_f|\psi_i}|^4}}=\frac{\Delta_f\hat{O}}{|\braket{\psi_f|\psi_i}|^2}\, ,
\end{equation}
which is proportional to the uncertainty of the operator $\hat{O}$ in the final state $\ket{\psi_f}$.

\subsection{Some analytical formulas for $\sigma_x$}
\label{appendix:some_analytical_formulas}
\begin{figure}
    \centering
\includegraphics[scale=0.35]{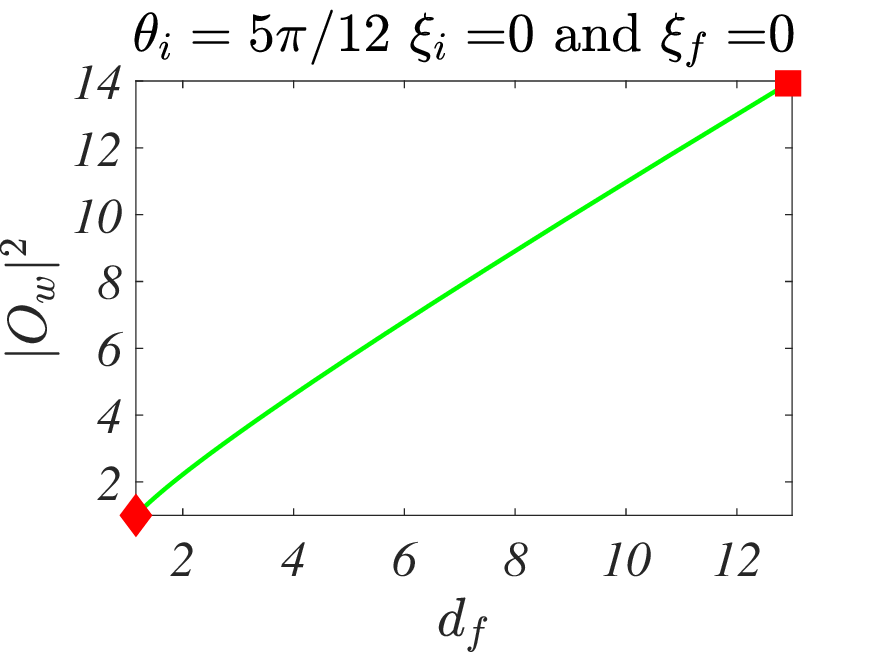}\quad\includegraphics[scale=0.27]{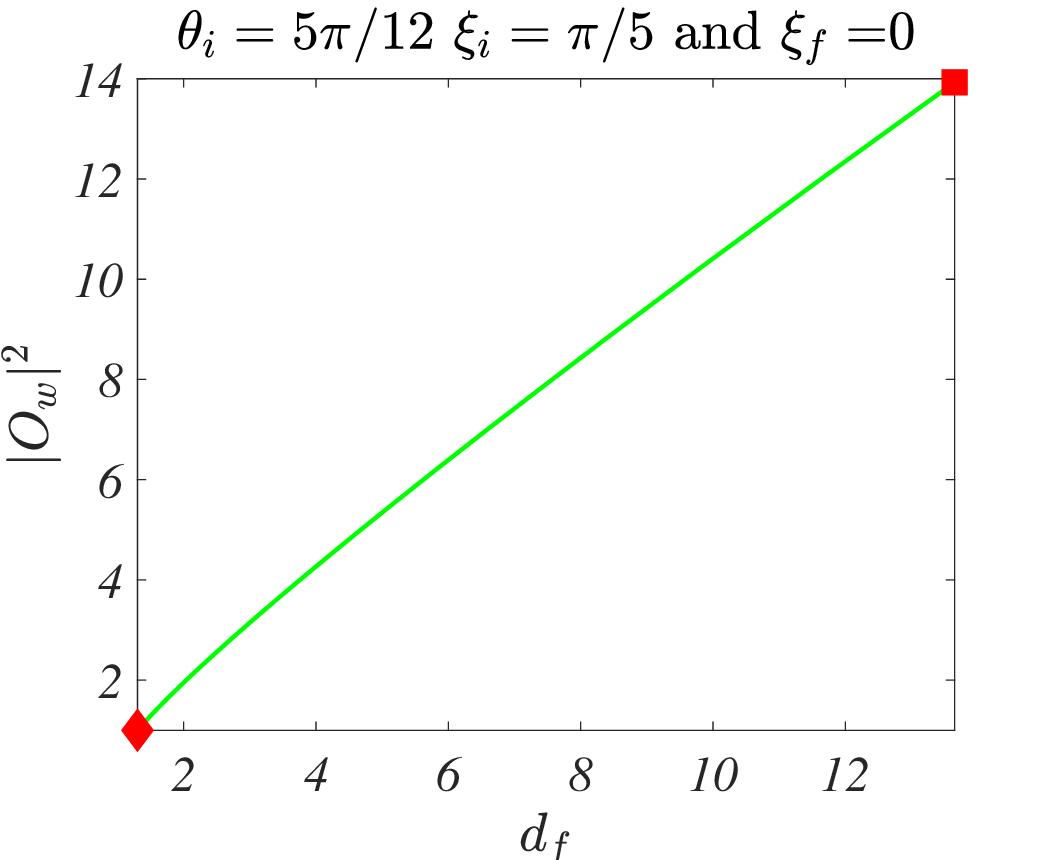}\quad\includegraphics[scale=0.35]{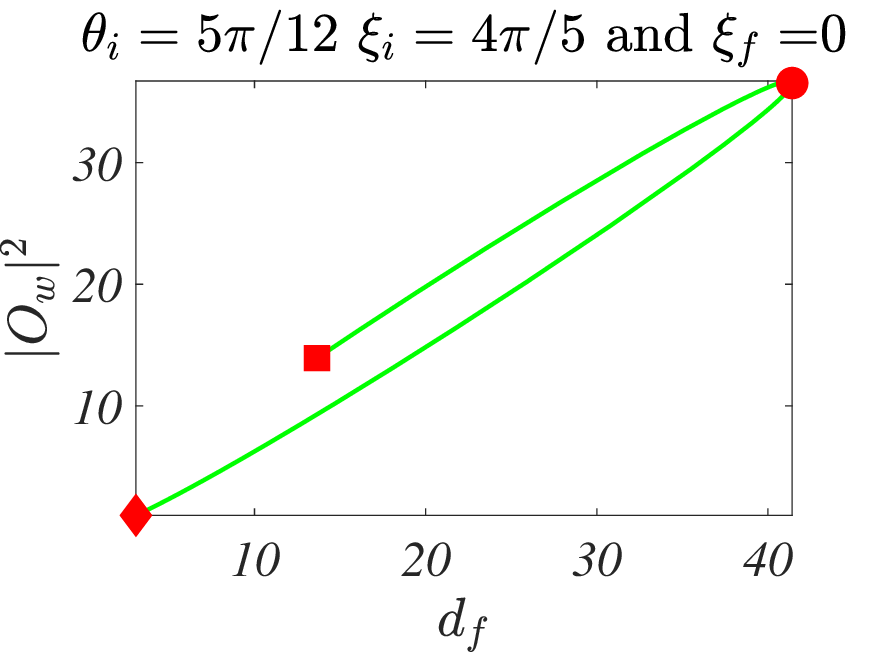}
\caption{We show the curves $d_f(\hat{A}_x)$ and $|\hat{\sigma}_{x,w}|^2$ as function of $\theta_f$ in the range $[0,\tilde{\theta}_f]$, where the upper bound is determined by the condition $|\hat{\sigma}_{x,w}|^2(\tilde{\theta}_f)=1$ and it is marked with a red diamond. The value $\theta_f=0$ is marked with a red square, while the value $\hat{\theta}_f$ at which $d_f$ attains its maximum is marked by a red circle, which denotes once the latter falls in the domain of anomalous weak value. The remaining parameters have been fixed to $\theta_i=5\pi/12$, $\xi_i=0$ and $\xi_f=0$ (left panel), $\theta_i=5\pi/12$, $\xi_i=\pi/5$ and $\xi_f=0$ (middle panel), $\theta_i=5\pi/12$, $\xi_i=4\pi/5$ and $\xi_f=0$ (right panel).}
    \label{fig:modOw2Vsdf}
\end{figure}

The aim of this section is to present the analytical expressions relating $O_w$ and $d_f$ for the cases numerically studied in the main text.
Let us assume the pre- and post-selected states to be given in the general form
\begin{eqnarray*}
\ket{\psi_a}=
\begin{pmatrix}
\cos{\theta_a}\\
e^{i\xi_a}\sin{\theta_a}
\end{pmatrix},
\end{eqnarray*}
where for $a\in\{i,f\}$ we assume $0\leq\theta_a\leq\frac{\pi}{2}$ and $0\leq\xi_a\leq 2\pi$. For sake of definitiveness, let us consider the case of the $x$-Pauli matrix. Because $\hat{\sigma}_x^2=1$ and the normalization of the initial state, we get from Eq.~\eqref{eq:dfA}
\begin{equation}
\label{eq:dfsigmax}
d_f\left(\hat{A}_x\right)=\frac{\sqrt{1 - \sin^2(2\theta_i)\cos^2\xi_i}}{|\braket{\psi_f|\psi_i}|^2}\, ,
\end{equation}
where
\begin{equation*}
|\braket{\psi_f|\psi_i}|^2 = \cos^2\theta_f\cos^2\theta_i+\sin^2\theta_f\sin^2\theta_i+2\cos(\xi_i-\xi_f)\cos\theta_f\cos\theta_i\sin\theta_f\sin\theta_i\, .
\end{equation*}
To compute the weak value we use the very first definition~\eqref{eq:weak_value_in_terms_of_trace} to obtain
\begin{equation}
\label{eq:modOwpow2}
|\sigma_{x,w}|^2=\frac{\sin^2\theta_f\cos^2\theta_i+\cos^2\theta_f\sin^2\theta_i+2\cos(\xi_i+\xi_f)\cos\theta_f\cos\theta_i\sin\theta_f\sin\theta_i}{|\braket{\psi_f|\psi_i}|^2}\, .
\end{equation}
Both $d_f\left(\hat{A}_x\right)$ and $|\sigma_{x,w}|^2$ depend on the angles, $\theta_f$ and $\theta_i$, and the phases $\xi_i$ and $\xi_f$. In the following, we will assume the last three quantities as fixed parameters and thus consider $X=d_f\left(\hat{A}_x\right)$ and $Y=|\sigma_{x,w}|^2$ to represent a curve parametrized by $\theta_f$ in the plane $(X,Y)$. 
In Fig.~\ref{fig:modOw2Vsdf}, we report three typical behaviors of such a curve. In the left panel, we assume $\xi_i=\xi_f=0$, and one clearly observes a positive correlation between $d_f(\hat{A}_x)$ and $|\hat{\sigma}_{x,w}|^2$: they both increases as $\theta_f$ varies from $\tilde{\theta}_f$, i.e., the values at which $|\hat{\sigma}_{x,w}|^2=1$ to $\theta_f=0$, corresponding here to the maximum of both $d_f(\hat{A}_x)$ and $|\hat{\sigma}_{x,w}|^2$. The results shown in the middle panel have been obtained by setting $\xi_i=\pi/5$ and still $\xi_f=0$ and we can observe the same behavior than the one presented in the left panel. A similar behavior persists once we increase $\xi_i$, up to a point at which a new behavior emerges (see right panel), at which the relation between $d_f(\hat{A}_x)$ and $|\hat{\sigma}_{x,w}|^2$ is no longer monotonic. In this panel, we can indeed observe the existence of a value $\hat{\theta}_f$ at which $d_f$ attains its maximum and this value is in the region of anomalous weak values.
The value $\tilde{\theta}_f$ can be found by setting equation~\eqref{eq:modOwpow2} equal to $1$ and solving the resulting relation for $\tan \theta_f$. A straightforward computation returns, if $\tan^2\theta_i\neq 1$,
\begin{equation*}
    \tan \tilde{\theta}_f = \tan (2\theta_i) \sin\xi_i\sin\xi_f \pm \sqrt{\tan^2 (2\theta_i) \sin^2\xi_i\sin^2\xi_f+1}\, ,
\end{equation*}
and we have to select the angle lying in $[0,\pi/2]$.

To determine the angle $\hat{\theta}_f$ that maximizes $d_f$, one can observe that the numerator of Eq.~\eqref{eq:dfsigmax} does not depend on $\theta_f$ and thus this is equivalent to minimizing the denominator. By doing so, we obtain
\begin{equation*}
\tan(2\hat{\theta}_f)=\tan(2\theta_i)\cos(\xi_i-\xi_f)\, ,
\end{equation*}
if $\cos(2\theta_i)\neq 0$ and $\hat{\theta}_f=\pi/4+k\pi/2$ if if $\cos(2\theta_i)= 0$.

Having those two angles, $\hat{\theta}_f$ and $\tilde{\theta}_i$, we can obtain conditions on $\xi_f$, $\xi_i$ and $\theta_i$ to determine if there is or not a monotonic dependence between $d_f(\hat{A}_x)$ and $|\hat{\sigma}_{x,w}|^2$.

To make some analytical progress we can explicitly express the dependence of $|\sigma_{x,w}|^2$ on $d_f(\hat{A}_x)$, again by removing $\theta_f$. To do so, we first express~\eqref{eq:dfsigmax} in terms of $\tan^2\theta_f$
\begin{equation*}
d_f\left(\hat{A}_x\right)=\sqrt{1 - \sin^2(2\theta_i)\cos^2\xi_i}\frac{(1+\tan^2\theta_i)(1+\tan^2\theta_f)}{1+\tan^2\theta_f\tan^2\theta_i+2\cos(\xi_i-\xi_f)\tan\theta_i\tan\theta_f}\, ,
\end{equation*}
and we solve the second degree equation in $\tan\theta_f$ to express the latter as a function of $d_f$ and the remaining variables
\begin{equation}
\tan\theta_f=\frac{d_f\cos(\xi_i-\xi_f)\tan\theta_i\pm\sqrt{d_f^2\cos^2(\xi_i-\xi_f)\tan^2\theta_i-[d_f\tan^2\theta_i-(1+\tan^2\theta_i) S][d_f-(1+\tan^2\theta_i)S]}}{d_f\tan^2\theta_i-(1+\tan^2\theta_i)S}\, ,
\label{eq:tanthetai}
\end{equation}
where $S=\sqrt{1-\sin^2(2\theta_i)\cos^2\xi_i}$.

We hence explicitly rewrite Eq.~\eqref{eq:modOwpow2} in terms of $\tan\theta_f$ 
\begin{equation*}
|\hat{\sigma}_{x,w}|^2=\frac{1}{1+\tan^2 \theta_i}\frac{1}{1+\tan^2 \theta_f}\frac{\tan^2 \theta_f+\tan^2\theta_i+2\cos(\xi_i+\xi_f)\tan\theta_f\tan\theta_i}{\sqrt{1-\sin^2(2\theta_i)\cos^2\xi_i}}d_f\, ,
\end{equation*}
and eventually use the expression~\eqref{eq:tanthetai} for $\tan\theta_f$.

\begin{figure}
    \centering
    \includegraphics[scale=0.38]{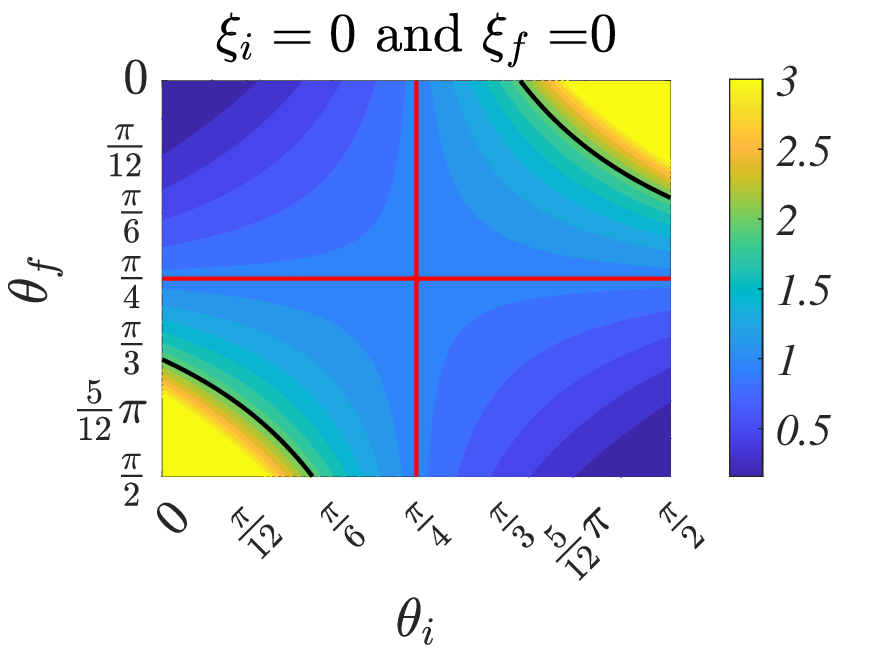}\quad\includegraphics[scale=0.38]{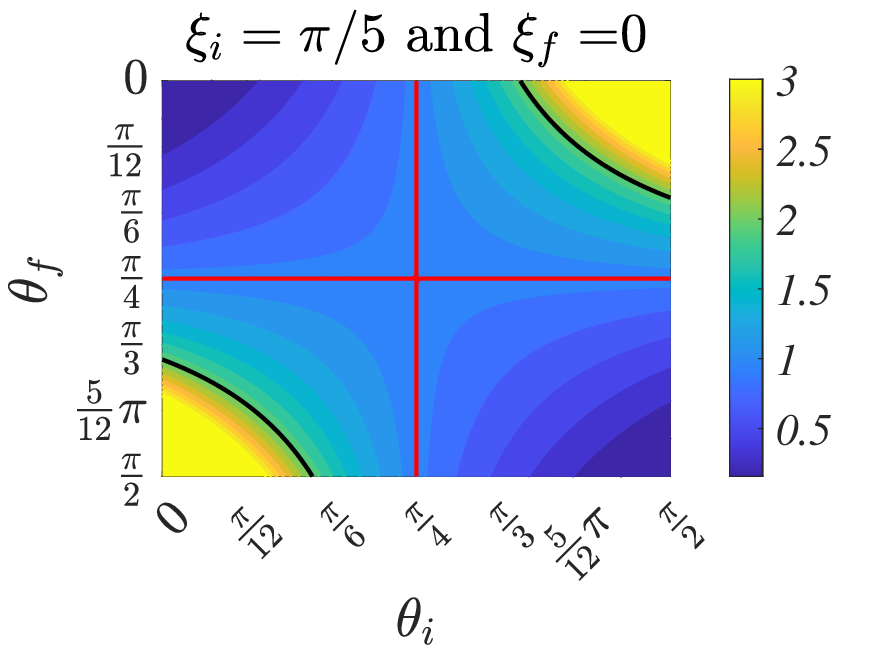}\quad\includegraphics[scale=0.38]{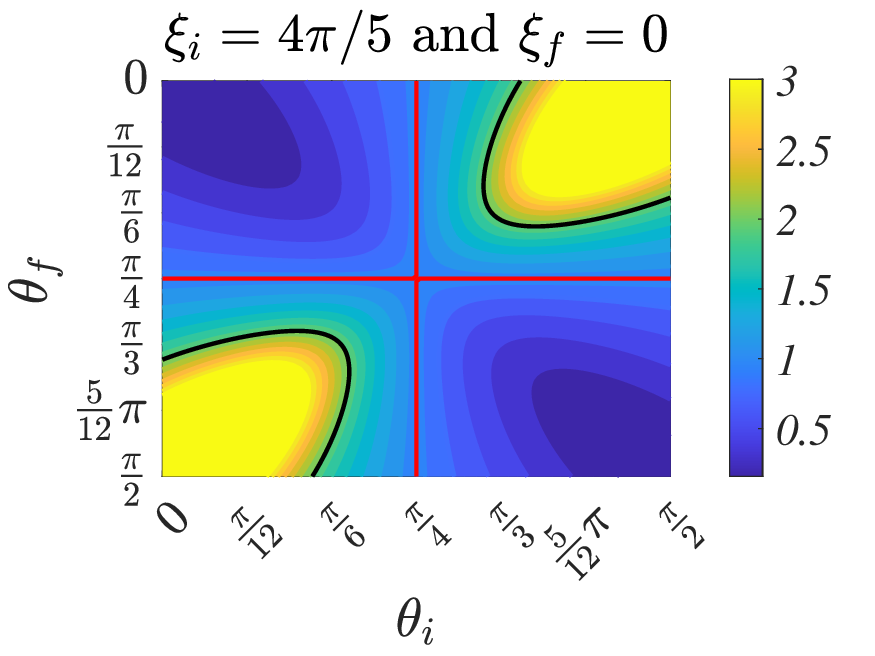}\\
    \includegraphics[scale=0.38]{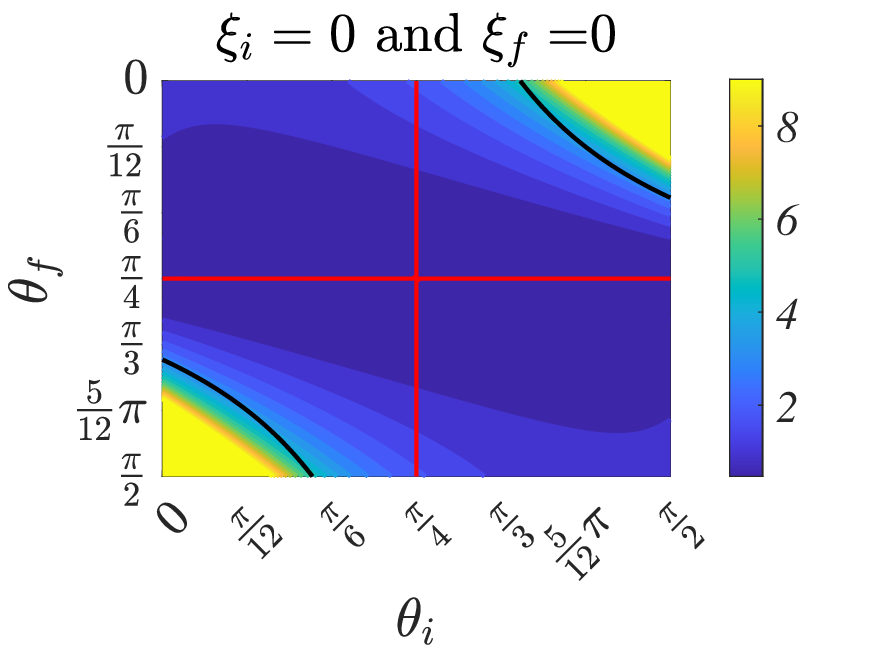}\quad\includegraphics[scale=0.38]{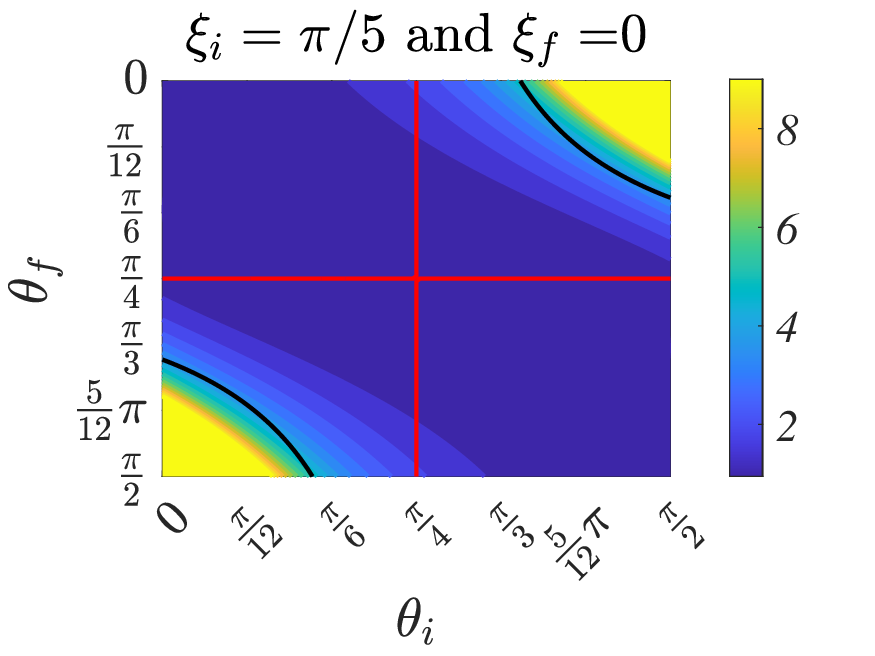}\quad\includegraphics[scale=0.38]{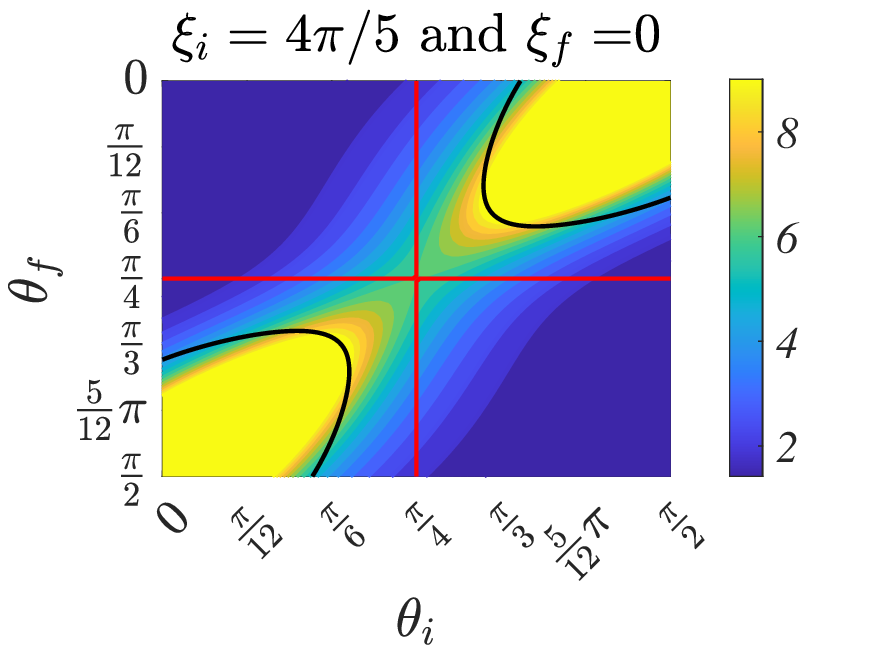}
    \caption{We show in the plane $(\theta_i,\theta_f)$ the level curves of the functions $|\hat{\sigma}_{x,w}|$ (top panels) and $d_f(\hat{A}_x)$ (bottom panels). The remaining parameters have been fixed to $(\xi_i,\xi_f)=(0,0)$ (left panel), $(\xi_i,\xi_f)=(\pi/5,0)$ (middle panel) and $(\xi_i,\xi_f)=(4\pi/5,0)$ (right panel).}
    \label{fig:figure_sigma_xTC}
\end{figure}

\begin{figure}
    \centering
\includegraphics[scale=0.38]{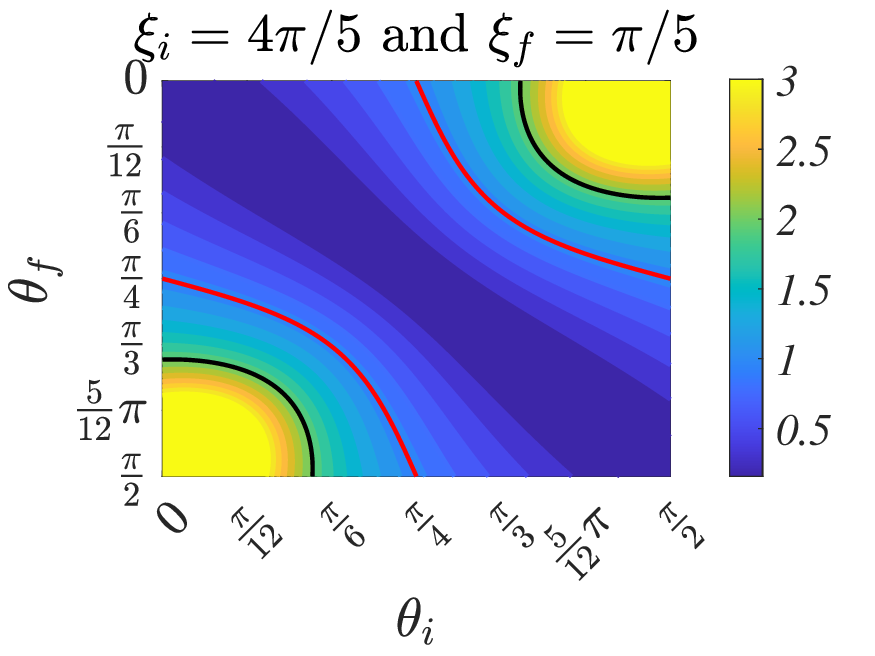}\quad\includegraphics[scale=0.38]{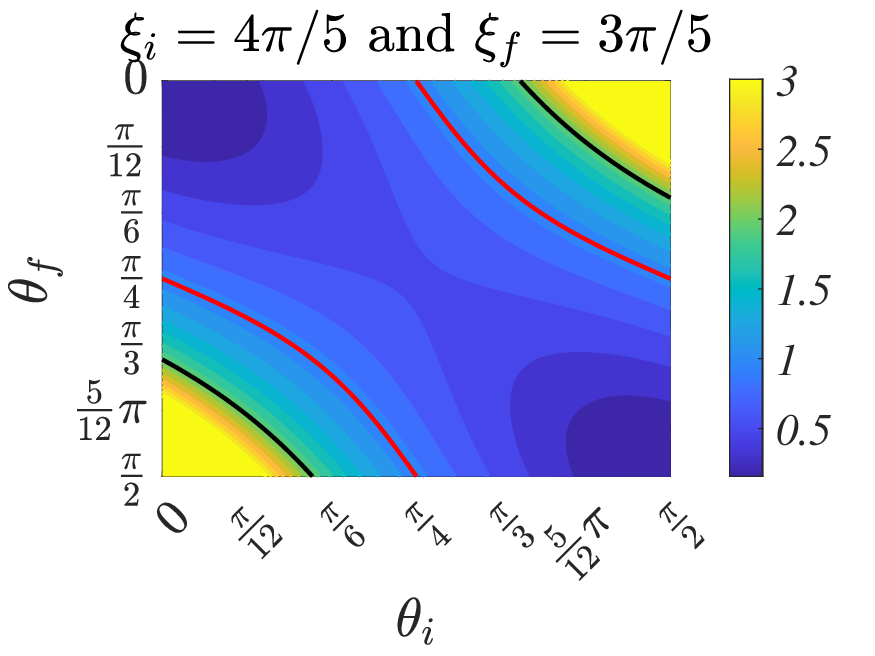}
    \caption{We show in the plane $(\theta_i,\theta_f)$ the level curves of the function $|\hat{\sigma}_{x,w}|$ for two sets of parameters $(\xi_i,\xi_f)=(4\pi/5,\pi/5)$ (left panel), $(\xi_i,\xi_f)=(4\pi/5,3\pi/5)$ (right panel).}
    \label{fig:figure_sigma_x2TC}
\end{figure}

We can observe that the curves given parametrically by Eqs.~\eqref{eq:dfsigmax} and~\eqref{eq:modOwpow2} shown in Fig.~\ref{fig:modOw2Vsdf} correspond to the branch of $\tan\theta_f$ Eq.~\eqref{eq:tanthetai} with the positive sign in front of the square root.
\begin{figure}
    \centering
\includegraphics[scale=0.35]{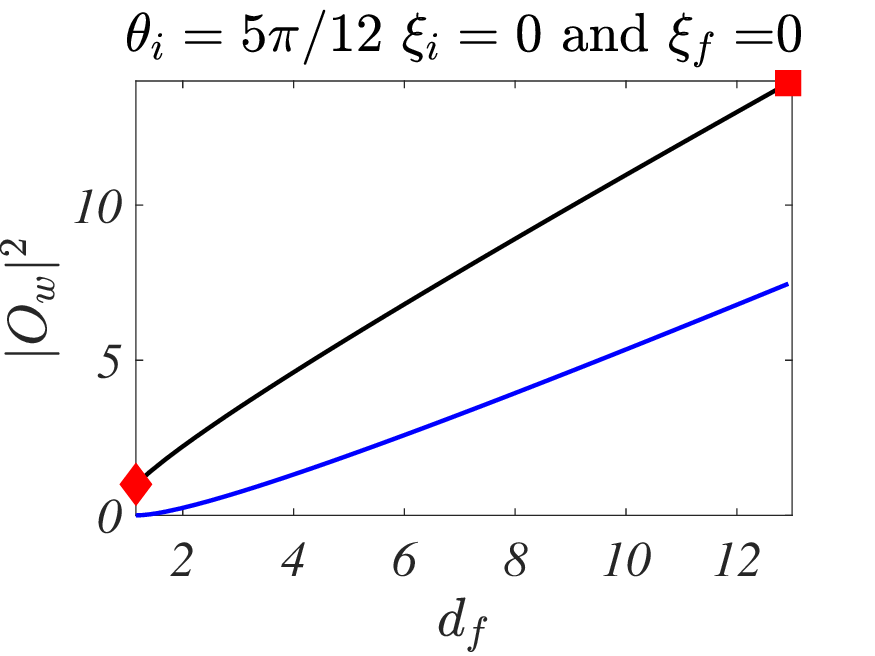}\quad\includegraphics[scale=0.35]{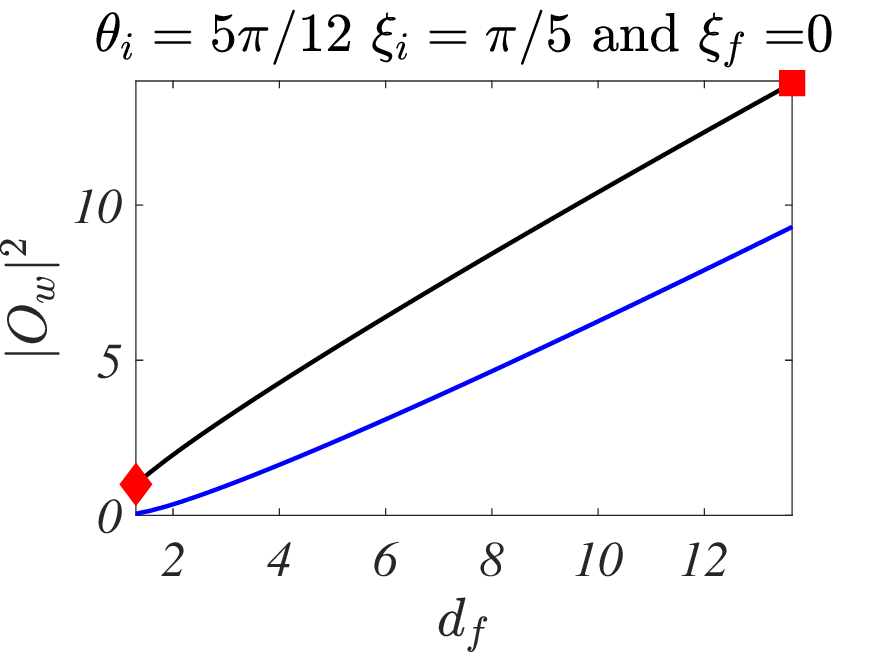}\quad\includegraphics[scale=0.35]{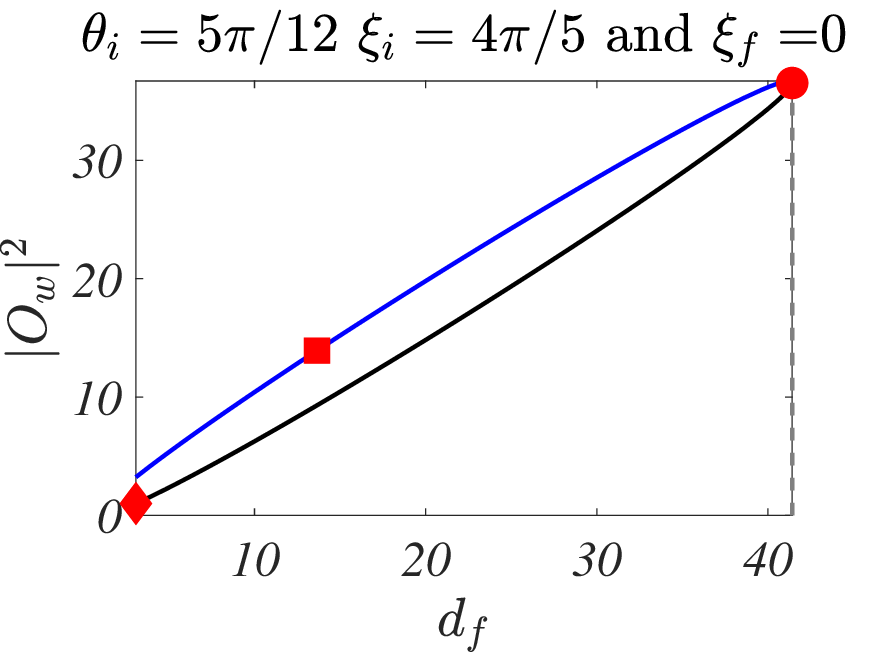}
\caption{We show the curve $|\hat{\sigma}_{x,w}|^2$ as a function of  $d_f(\hat{A}_x)$, in the range of weak value amplification. The black line corresponds to the root with the positive sign while the blue one to the negative sign. The red diamond marks the value of $d_f$ for which $|\hat{\sigma}_{x,w}|^2=1$. The red square identifies the value of $d_f$ associated to $\theta_f=0$, while the red circle determines the largest value of $d_f$. The remaining parameters have been fixed to $\theta_i=5\pi/12$, $\xi_i=0$ and $\xi_f=0$ (left panel), $\theta_i=5\pi/12$, $\xi_i=\pi/5$ and $\xi_f=0$ (middle panel), $\theta_i=5\pi/12$, $\xi_i=4\pi/5$ and $\xi_f=0$ (right panel).}
\label{fig:modOw2VsdfBranches}
\end{figure}
In Fig.~\ref{fig:figure_sigma_xTC} we propose a global view of $|\sigma_{x,w}|$ (top panels) and $d_f(\hat{A}_x)$ (bottom panels) as a function of $\theta_i$ and $\theta_f$ for three sets of values of $(\xi_i,\xi_f)$, $(0,0)$ (left panel), $(\pi/5,0)$ (middle panel) and $(4\pi/5,0)$ (right panel). We can observe that as $\xi_i$ increases, so does the region corresponding to strong anomalous weak values, e.g., larger than $2$.
On the other hand (see Fig.~\ref{fig:figure_sigma_x2TC}), increasing $\xi_f$ reduces the region of anomalous weak values.
\section{Analytical calculations with null phases}\label{appendix:analytical_calculations_considering_null_phases}
\begin{figure}
    \centering
    \includegraphics[scale=0.36]{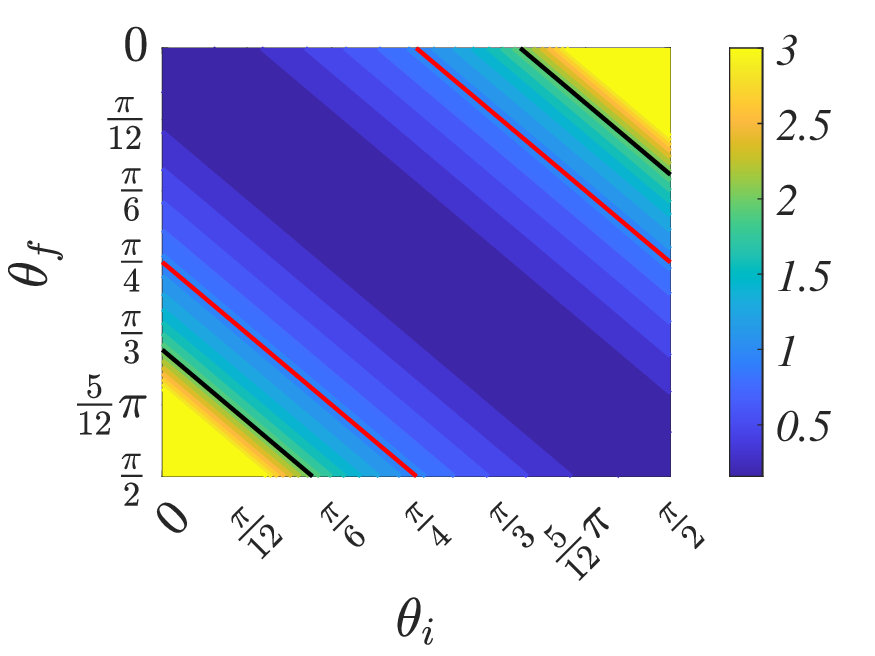}\quad\includegraphics[scale=0.36]{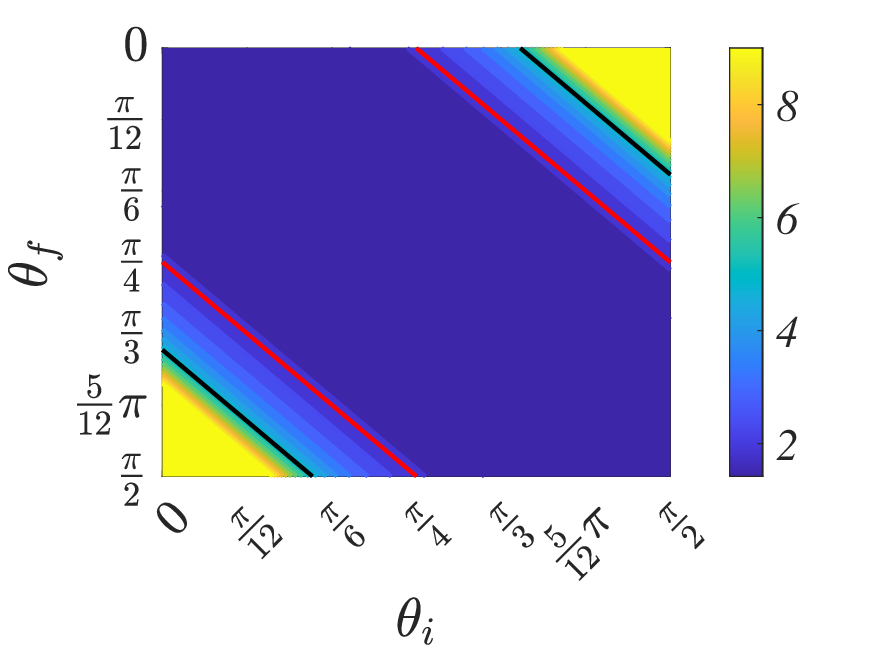}\quad\includegraphics[scale=0.36]{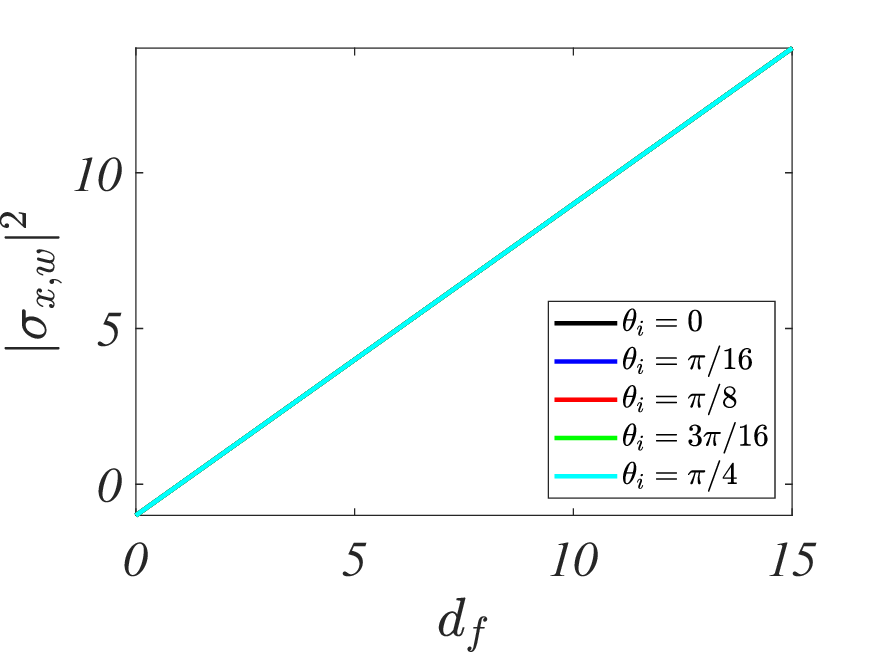}
    \caption{Henrici departure from normality of $\hat{A}_y$ in terms of $\theta_i$ and $\theta_f$, weak value of $\hat{\sigma}_y$ in terms of $\theta_i$ and $\theta_f$, and weak value of $\hat{\sigma}_y$ in terms of the Henrici departure in the anomalous regime from normality for different values of $\theta_i$ from $0$ to $\frac{\pi}{4}$.}
    \label{fig:figure_sigma_yTC}
\end{figure}

\begin{figure}
    \centering
    \includegraphics[scale=0.36]{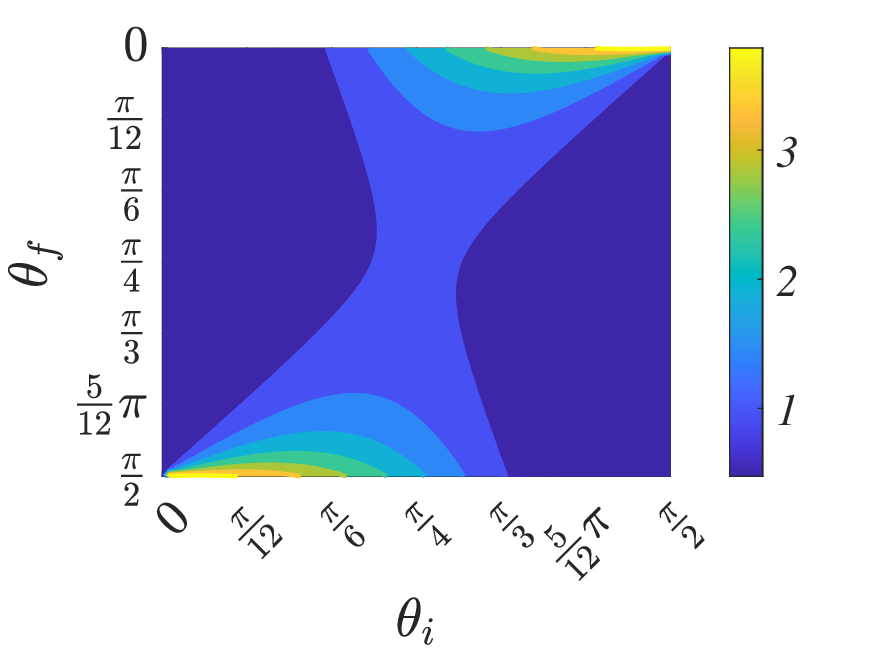}\quad\includegraphics[scale=0.36]{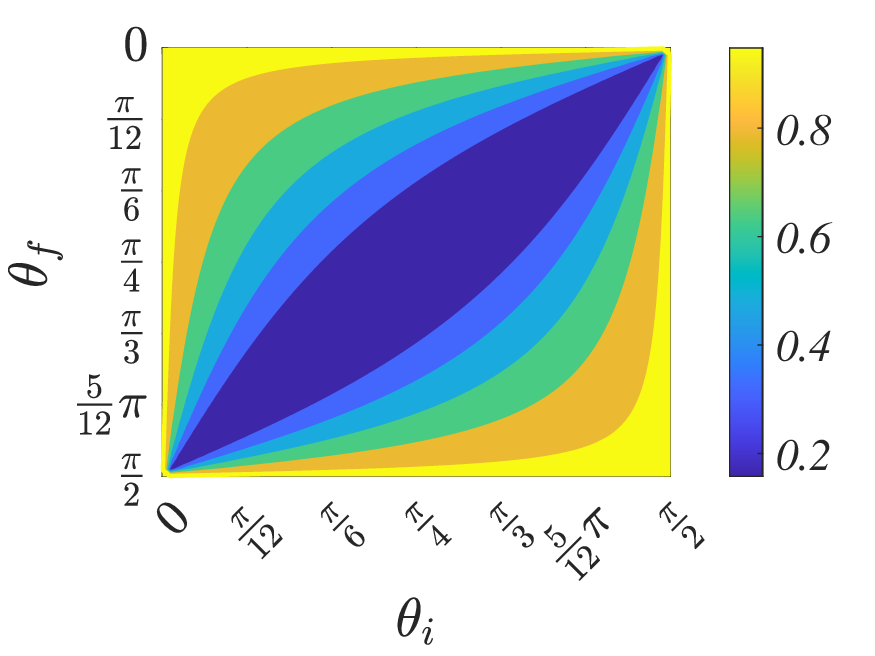}\quad\includegraphics[scale=0.27]{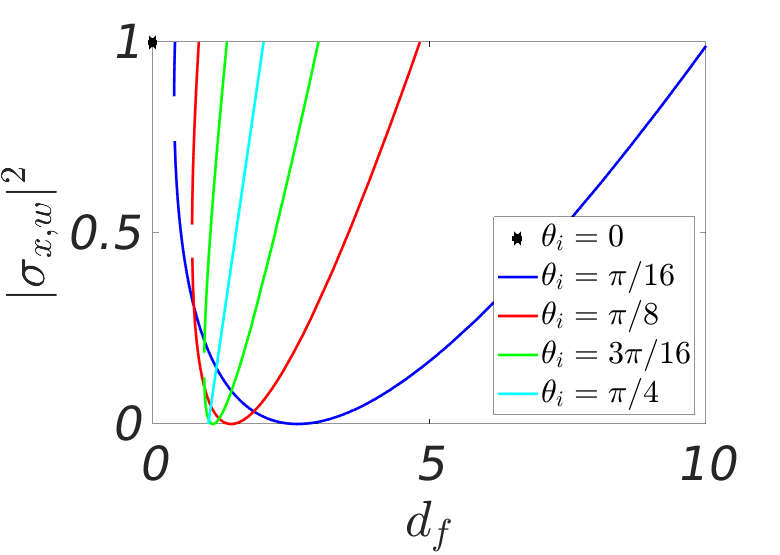}
    \caption{Henrici departure from normality of $\hat{A}_z$ in terms of $\theta_i$ and $\theta_f$, weak value of $\hat{\sigma}_z$ in terms of $\theta_i$ and $\theta_f$, and weak value of $\hat{\sigma}_z$ in terms of the Henrici departure from normality for different values of $\theta_i$ from $0$ to $\frac{\pi}{4}$.}
    \label{fig:figure_sigma_zTC}
\end{figure}

Let us consider now the Pauli matrix $\hat{\sigma}_y$. In this case, the Henrici departure from normality of the matrix $\hat{A}$ is, assuming $\xi_i=\xi_f=0$, 
\begin{equation}
    df\left(\hat{A}_y\right)=\frac{1}{\cos^2\left(\theta_f-\theta_i\right)},
\end{equation}
while the weak value is 
\begin{equation}
    \sigma_{y,w}=i\tan{\left( \theta_f-\theta_i\right)}.
\end{equation}
The Henrici departure from normality can be expressed in terms of the weak value as
\begin{equation}
    |\sigma_{y,w}|^2=\frac{1}{\cos^2\left(\theta_f-\theta_i\right)}-1=df\left(\hat{A}_y\right)-1.
\end{equation}
A comparison between Fig.\ref{fig:figure_sigma_yTC} and Fig.\ref{fig:figure_sigma_y} reveals a remarkable agreement between the figures generated from the numerical and analytical studies. 
Finally, we also studied the Pauli matrix $\hat{\sigma}_z$, assuming $\xi_i=\xi_f=0$. The Henrici departure from normality is in that case, 
\begin{equation}
    d_f\left(\hat{A}_z\right)=\sin^2{2\theta_i}\left[ 1+\tan^2{\left(\theta_f-\theta_i\right)}\right],
\end{equation}
while the weak value is 
\begin{equation}
    |\sigma_{z,w}|=|\cos{2\theta_i}-\tan{\left(\theta_f-\theta_i\right)}\sin{2\theta_i}|.
\end{equation}
The weak value can be expressed in terms of the Henrici departure from normality as, 
\begin{equation}
    |\sigma_{z,w}|^2=\Bigg|\cos{2\theta_i}\pm\sin{2\theta_i}\sqrt{-1\pm\frac{df\left(\hat{A}_z\right)}{\sin{2\theta_i}}}\Bigg|^2.
\end{equation}
Figures~\ref{fig:figure_sigma_zTC} and \ref{fig:figure_sigma_z} depict the results of the theoretical and numerical calculations, and a striking similarity between the two figures is immediately apparent.

\section{Comparison between weak value and non-normality of the two different formulations of the operator}\label{appendix:sec4_new}
\begin{figure}
    \centering
\includegraphics[scale=0.35]{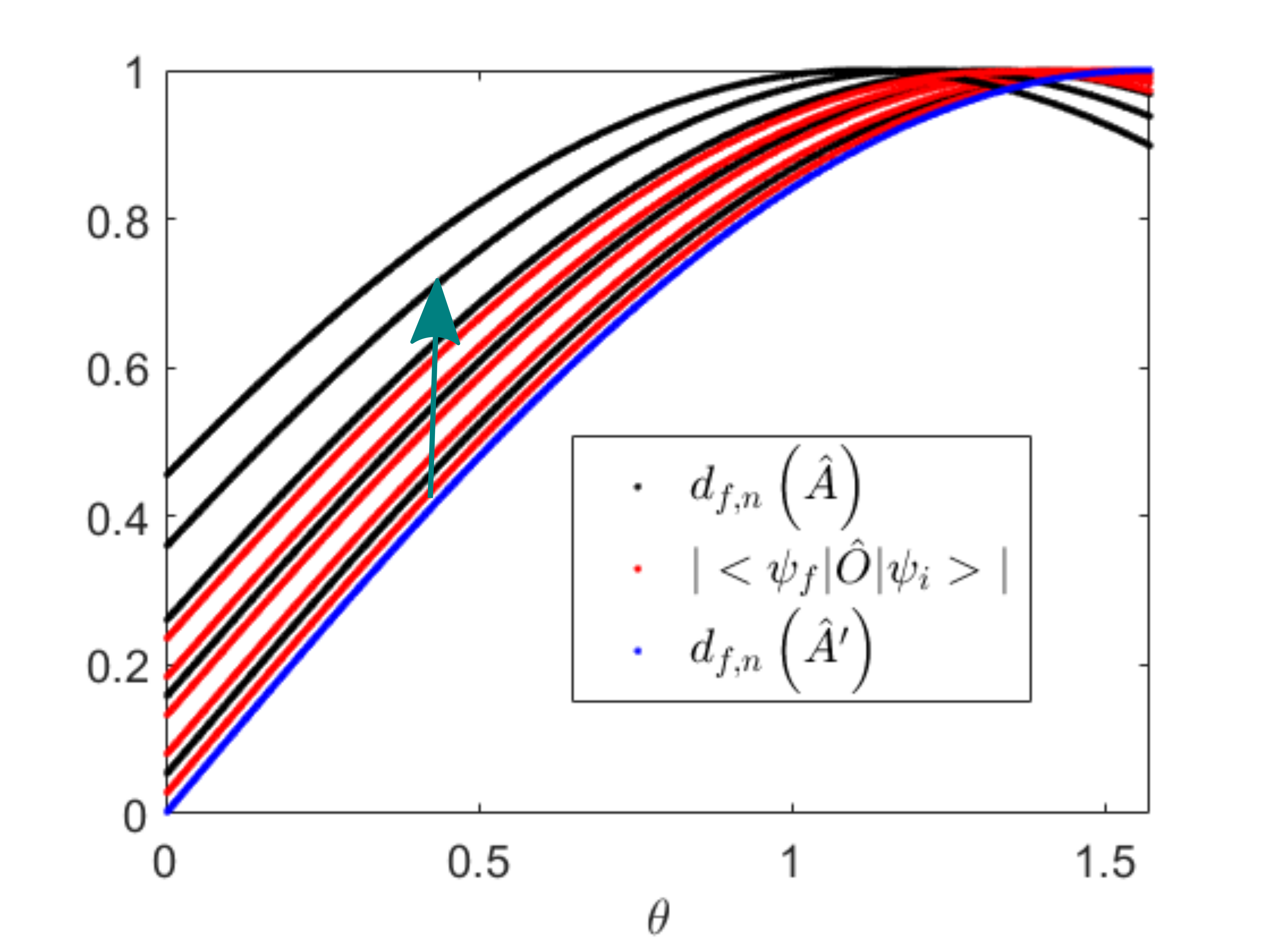}(a)\quad\includegraphics[scale=0.44]{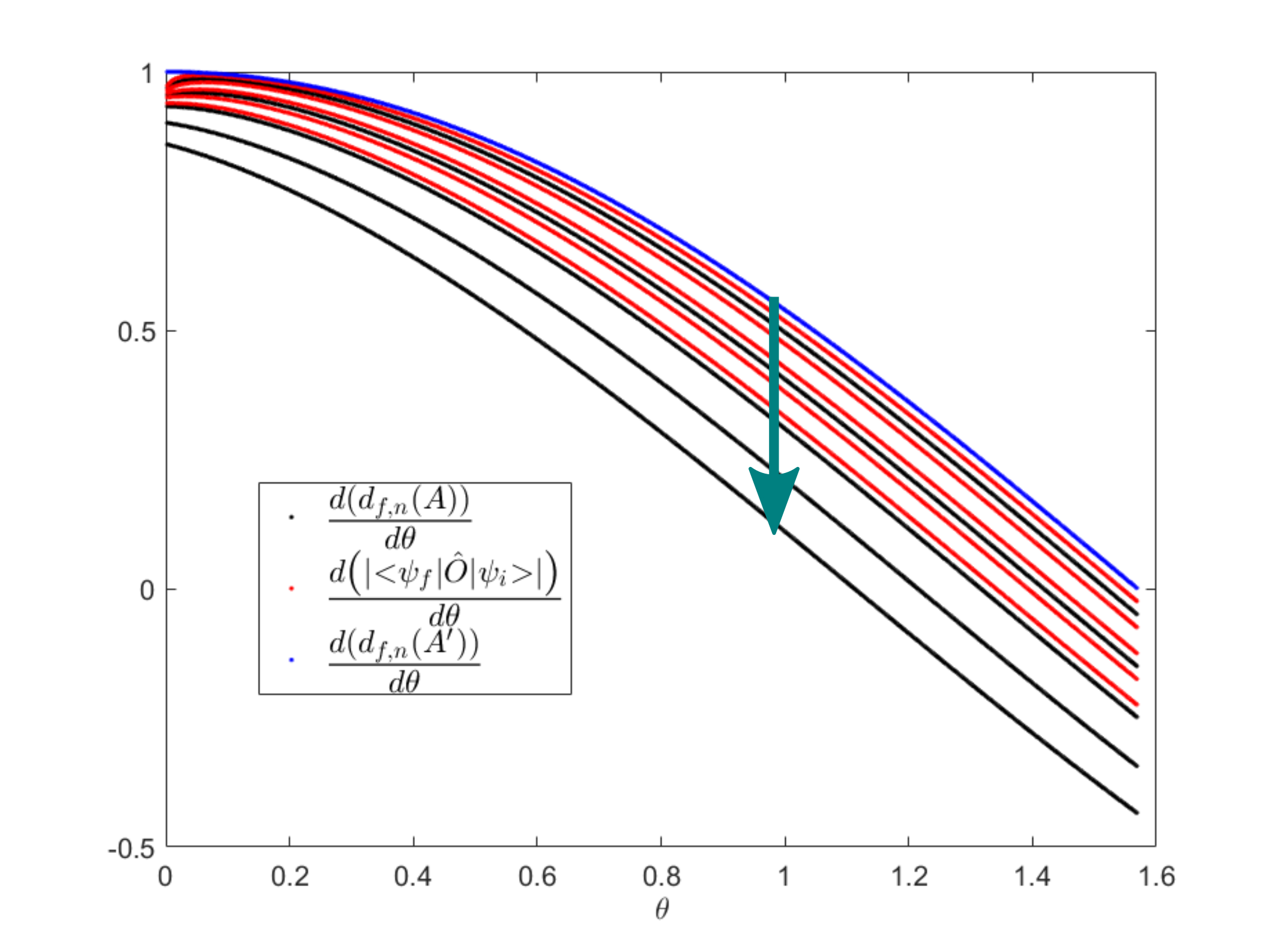}(b)
\caption{a) For different initial states ($\theta_i$), modulus of the numerator of the weak value (red), normalized Henrici departure from normality of $\hat{A}'$ (blue), and normalized Henrici departure from normality of $\hat{A}$ (black), all as a function of $\theta$ in the range of weak value amplification. b) For different initial states ($\theta_i$), derivative of the modulus of the numerator of the weak value $|\braket{\psi_f|\hat{O}|\psi_i}|$ (red), the normalized Henrici departure from normality of $\hat{A}'$ (blue), and the normalized Henrici departure from normality of $\hat{A}$ (black), all as a function of $\theta$ in the range of weak value amplification. The chosen parameters are: $\theta_f=\xi_i=\xi_f=0$, $\phi=\frac{\pi}{12}$, while $\theta_i$ varies from from $1.5446$ to $ 1.3352$. $\theta_i$ decreases in the direction of the green arrow.}
    \label{fig:figure_varying_operator_first_case}
\end{figure}

In this Appendix, we further develop the analysis of Section \ref{new_expression}, i.e., by examining how varying the observable impacts non-normality and the modulus of weak values, for given pre- and post selected states, for two-level systems. For sake of clarity, let us rewrite Eq. (\ref{eq:varying_operator}) from the main text, namely, that of an observable depending on two parameters,
\begin{equation}
\label{eq:varying_operator_app}
    \hat{O}=\sin{\theta}\cos{\phi}\ \hat{\sigma}_x+\sin{\theta}\sin{\phi}\ \hat{\sigma}_y+\cos{\theta}\ \hat{\sigma}_z\, , 
\end{equation}
where $0\leq \theta \leq \frac{\pi}{2}$, $0\leq \phi \leq 2\pi$, and $\hat{\sigma}_i$ are the Pauli matrices (see Appendix \ref{appendix:Pauli_matrices}). As previously stated, the goal is to explicit the relationship between the modulus of the numerator of the weak value, i.e., $|\bra{\psi_f}\hat{O}\ket{\psi_i}|$, and the normalized Henrici departure from normality for the two matrices, i.e., $d_{f,n}\left(\hat{A}\right)$ and $d_{f,n}\left(\hat{A}'\right)$. The results reported in Fig.\ref{fig:figure_varying_operator_first_case}a) show a clear trend. Since the operator $\hat{A}^\prime$ does not depend on the initial state, the normalized Henrici departure from normality $d_{f,n}\left(\hat{A}'\right)$ does not vary when we change the pre-selected state, with a fixed final state $\theta_f=0$; obviously, this is not the case for  $d_{f,n}\left(\hat{A}\right)$, as the different black curves clearly show. When the initial state is completely orthogonal to the final one, i.e., $\theta_i= \frac{\pi}{2}$, the two Henrici indexes and the weak value coincide. Starting from this value, the two indexes differ from each other but evolve similarly, and the weak value also follows a similar trend in between both departures from normality, in most of the range of values of $\theta$.
In order to understand this behavior, we study the derivative of the three functions (see Fig.~\ref{fig:figure_varying_operator_first_case}b)). As one can appreciate, the derivative of the weak value with respect to $\theta$ is always between the ones of the two normalized Henrici departures from normality, for the studied system. Consequently, the variation of the weak value is not only determined by the orthogonality of the pre- and post-selected states, as seen when varying $\theta_i$, but also by the normalized non-normality of the operators $\hat{A}$ and $\hat{A}'$. 
\begin{figure}
    \centering
   \includegraphics[scale=0.44]{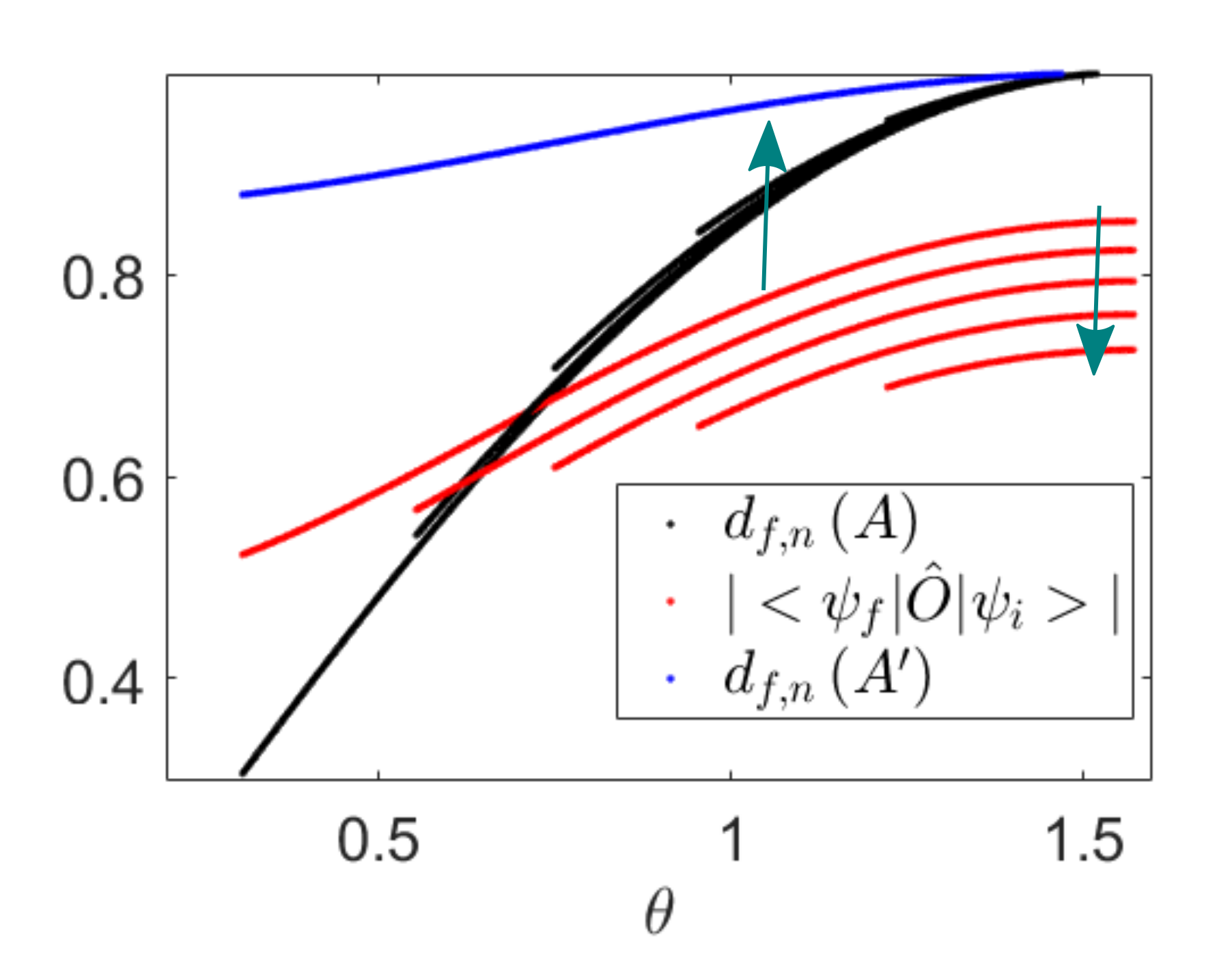}(a)\quad\includegraphics[scale=0.44]{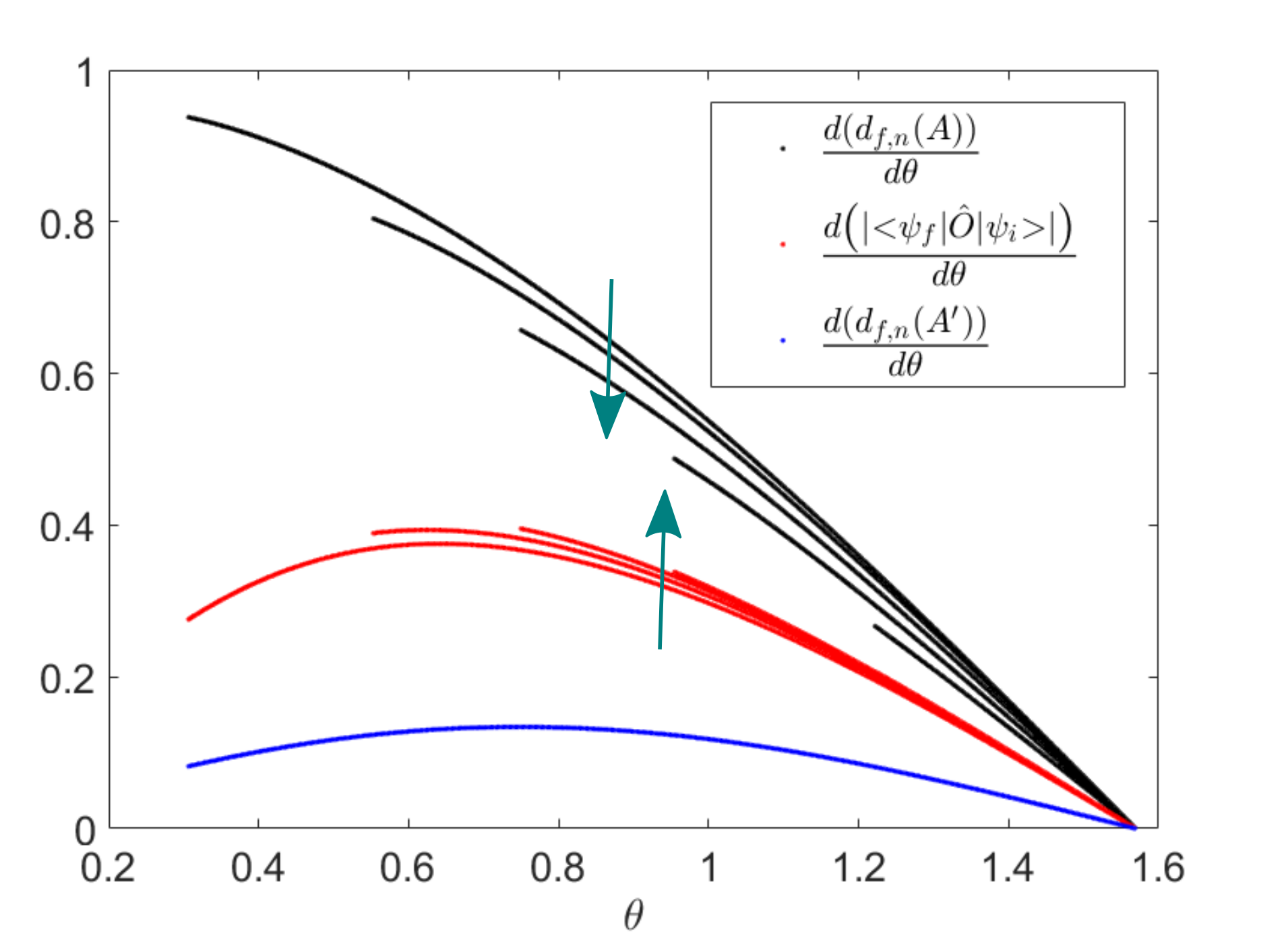}(b)
   \caption{a) For different initial states ($\theta_i$), modulus of the numerator of the weak value (red), normalized Henrici departure from normality of $\hat{A}'$ (blue), and normalized Henrici departure from normality of $\hat{A}$ (black), all as a function of $\theta$ in the range of weak value amplification.  b) For different initial states ($\theta_i$), derivative of the numerator of the modulus of the weak value $|\braket{\psi_f|\hat{O}|\psi_i}|$ (red), the normalized Henrici departure from normality of $\hat{A}'$ (blue), and the normalized Henrici departure from normality of $\hat{A}$ (black), all as a function of $\theta$ in the range of weak value amplification. The chosen parameters are: $\theta_f=\xi_i=\xi_f=0$, $\phi=3\frac{\pi}{2}$, while $\theta_i$ varies from from $1.5446$ to $ 1.3352$. $\theta_i$ decreases in the direction of the green arrow.}
    \label{fig:figure_varying_operator_second_case}
\end{figure}
In some cases, the behavior of numerator of the weak value and the normalized Henrici departures from normality can be significantly different, as shown in Fig.~\ref{fig:figure_varying_operator_second_case}. For the chosen parameters in this scenario, the numerator of the weak value does not fall between the two normalized Henrici departures from normality of $\hat{A}$ and $\hat{A}'$. Moreover, as $\theta_i$ decreases, the curve of the weak value moves down in the plot, whereas the normalized Henrici departure from normality of $\hat{A}$ moves in the opposite direction (see opposite orientations of the green arrows). It is worth noting that this type of behavior commonly occurs near the boundary of the amplification region of the weak values. Fig.~\ref{fig:weak_value_first_and_second_case} shows the modulus of the weak value plotted against $\theta$. In the second case, the modulus of the weak value does not even reach twice the largest eigenvalue of the observable, whereas in the first case, the anomalous weak value is significantly larger. 
\begin{figure}
    \centering
\includegraphics[scale=0.42]{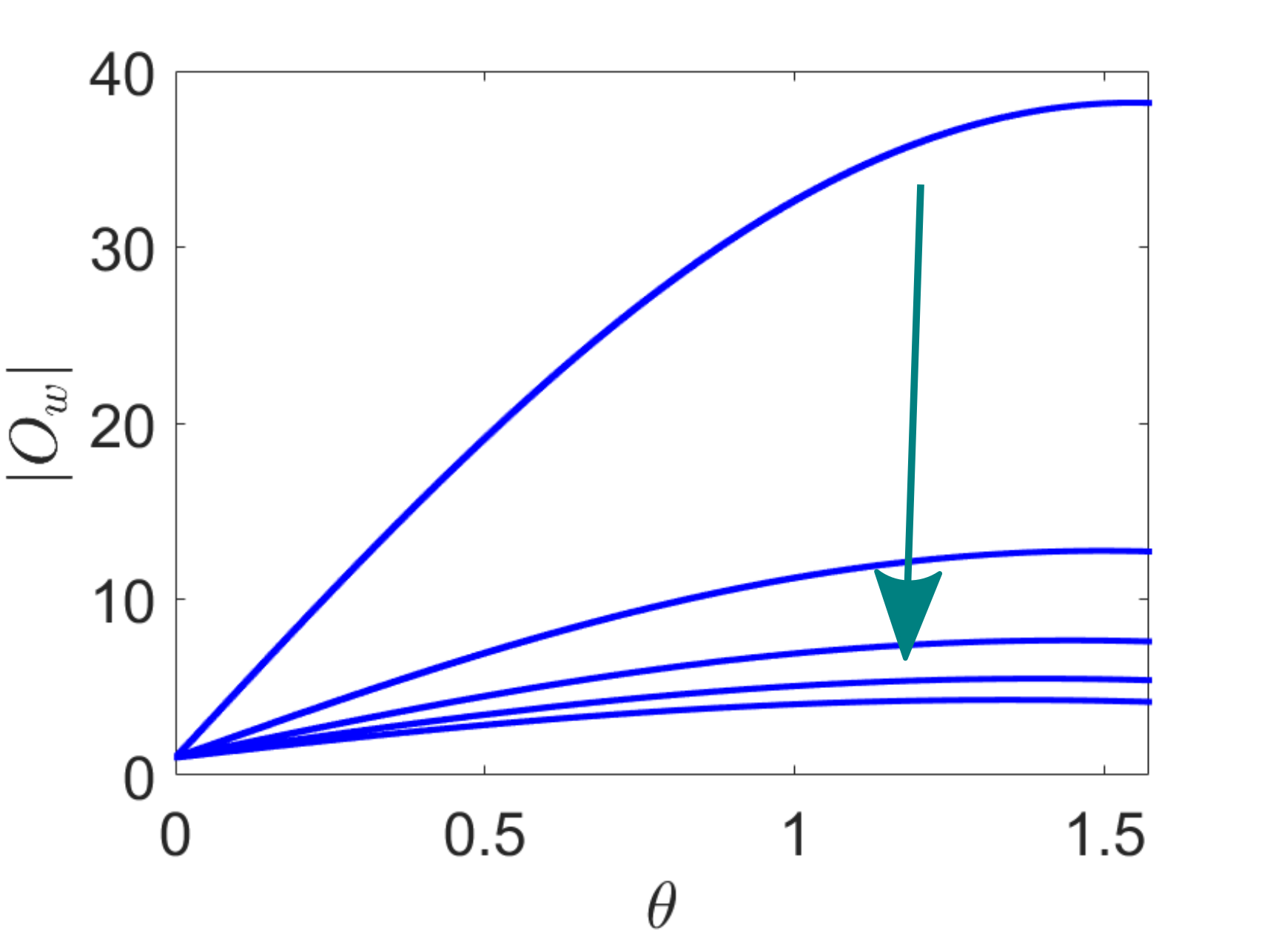}(a)\quad\includegraphics[scale=0.37]{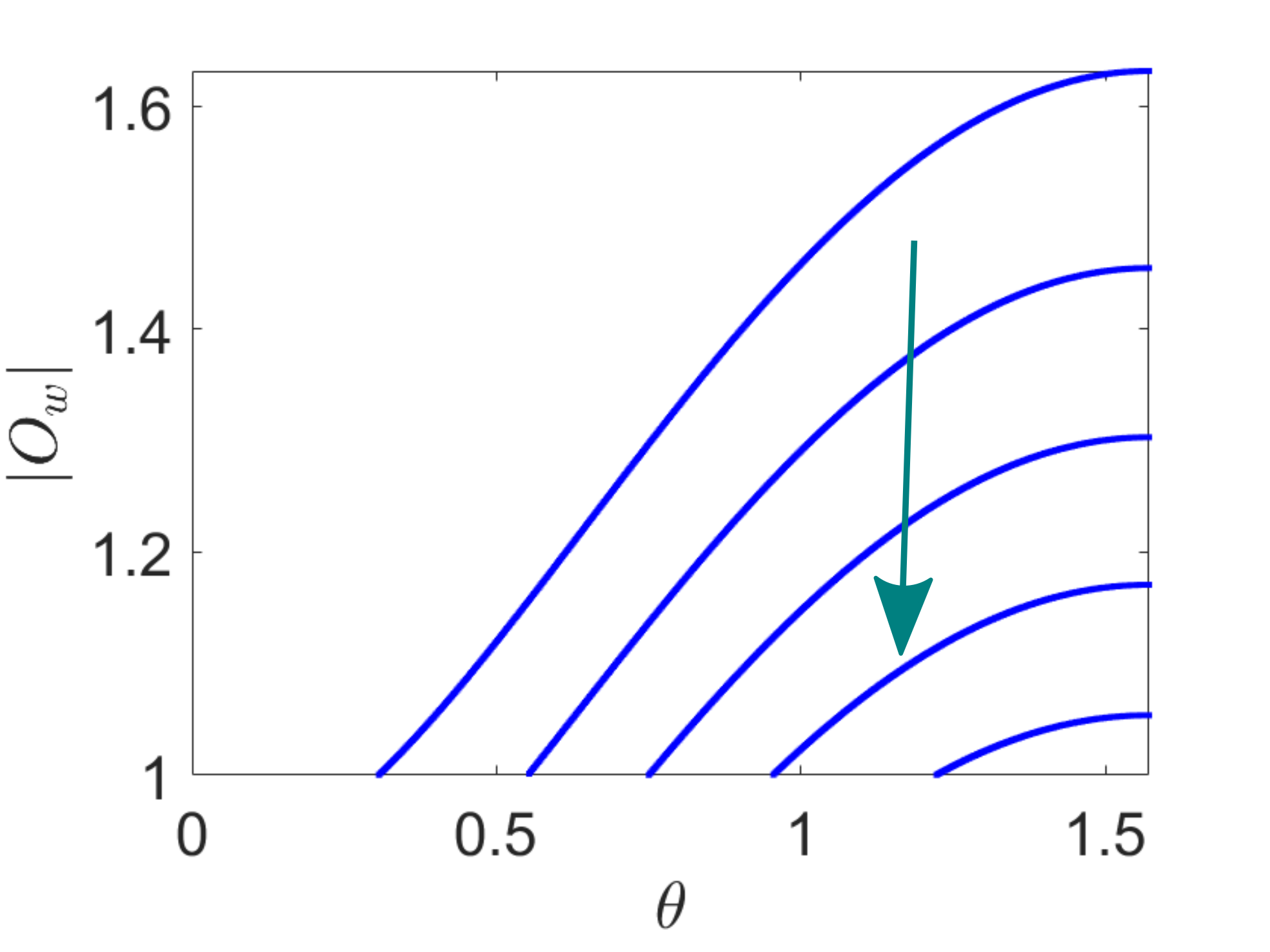}(b)
\caption{a) Modulus of the weak value of $\hat{O}$ in terms of $\theta$, for the cases of Fig. \ref{fig:figure_varying_operator_first_case}. The chosen parameters are: $\theta_f=\xi_i=\xi_f=0$, $\phi=\frac{\pi}{12}$, while $\theta_i$ varies from from $1.5446$ to $ 1.3352$. b) Modulus of the weak value of $\hat{O}$ in terms of $\theta$, for the cases of Fig. \ref{fig:figure_varying_operator_second_case}. The chosen parameters are: $\theta_f=\xi_i=\xi_f=0$, $\phi=3\frac{\pi}{2}$, while $\theta_i$ varies from from $1.5446$ to $ 1.3352$. $\theta_i$ decreases in the direction of the green arrow.}
    \label{fig:weak_value_first_and_second_case}
\end{figure}
The intricate relationship between the weak value and the Henrici departure from normality of $\hat{A}$ and $\hat{A}'$ in two-level systems is an area that warrants further investigation. Detailed studies can uncover more about this behavior. The interested readers can find the expressions of the weak value, the Henrici departure from normality, and their derivatives in Appendix~\ref{appendix:derivative}, with the assumption that $\xi_i=\xi_f=0$.

\section{Study of the point of maximum weak value and maximum Henrici departure from normality}\label{appendix:maximum_weak_value_maximum_df}
In this section, we will show some analytical results about the value of the observable parameter ($\theta$) in which the weak value presents a maximum and the ones in which both matrices ($\hat{A}$ and $\hat{A}'$) present a Henrici departure from normality that is the largest. 
Let us consider, without loss of generality, the specific example of Fig.\ref{fig:angle_nilpotent_angle}. The studied observable is
\begin{equation}
\hat{O}=
    \begin{pmatrix}
        \cos{\theta} & \frac{\left(1-i\right)}{\sqrt{2}}\sin{\theta} \\
        \frac{\left(1+i\right)}{\sqrt{2}}\sin{\theta} & -\cos{\theta}
    \end{pmatrix}.
\end{equation}
The pre- and post-selected states are
\begin{equation}
    \ket{\psi_i}=\begin{pmatrix}
        \cos{\theta_i}\\
        \sin{\theta_i}
    \end{pmatrix},
    \hspace{1 cm}
    \ket{\psi_f}=\begin{pmatrix}
        1\\
        0
    \end{pmatrix}.
\end{equation}
Both matrices $\hat{A}$ and $\hat{A}'$ have an eigenvalue that is equal to $0$. The other eigenvalues are
\begin{eqnarray}
    \alpha_{\hat{A}}&=&\cos{\theta},\\ \nonumber
    \alpha_{\hat{A}'}&=&\cos{\theta}\cos{2\theta_i}+\sqrt{2}\cos{\theta_i}\sin{\theta}\sin{\theta_i}.
\end{eqnarray}
The Henrici departures from normality of $\hat{A}$, and $\hat{A}'$ are
\begin{eqnarray}
        d_f\left(\hat{A}\right)&=&|\sin{\theta}| \\ \nonumber
    d_f\left(\hat{A}'\right)&=&\frac{1}{\sqrt{8}}\sqrt{\left(5-\cos{4\theta_i}-\cos{2\theta}\left(1+3\cos{4\theta_i}\right)-2\sqrt{2}\sin{2\theta}\sin{4\theta_i} \right)}.
\end{eqnarray}
The weak value is
\begin{equation}
    |O_w|=\sqrt{\cos^2{\theta}+\sqrt{2}\cos{\theta}\sin{\theta}\tan{\theta_i}+\sin^2{\theta}\tan^2{\theta_i}}.
\end{equation}
At $\theta=0$, the weak value is equal to $1$, independently on the value of the initial polar angle, $\theta_i$. When $0\leq\theta\leq\frac{\pi}{4}$, the weak value is only anomalous for a section of values of $\theta$, actually, with $0\leq\theta\leq\arctan{\frac{\sqrt{2}\tan{\theta_i}}{1-\tan^2{\theta_i}}}$. 

The Henrici departure from normality is the largest when the eigenvalue $\alpha$ is the smallest in absolute value. In the case of $\alpha_{\hat{A}}$, it always decreases in the range of $\theta$ between $0$ and $\frac{\pi}{2}$. Consequently, the maximum value of the Henrici departure from normality in the range of anomalous weak values would be at $\theta=\arctan{\frac{\sqrt{2}\tan{\theta_i}}{1-\tan^2{\theta_i}}}$. 

The eigenvalue of $\hat{A}'$ that is different from zero, $\alpha_{\hat{A}'}$ increases in absolute value when increasing $\theta$ in the range $0\leq\theta\leq\arctan{\frac{\sqrt{2}\tan{\theta_i}}{1-\tan^2{\theta_i}}}$. The point of the largest Henrici departure from normality in the anomalous regime is at $\theta=0$, for all values of $0\leq\theta_i\leq\frac{\pi}{4}$.

The value of $\theta$ for which the weak value is maximum is exactly at the average of both points of the largest value of Henrici departure from normality, $\theta=\frac{1}{2}\arctan{\frac{\sqrt{2}\tan{\theta_i}}{1-\tan^2{\theta_i}}}$.

When moving to $\frac{\pi}{4}\leq\theta_i\leq\frac{\pi}{2}$, the weak value is anomalous for all values of $\theta$ in the range $0\leq\theta\leq\frac{\pi}{2}$. The largest Henrici departure from normality of $\hat{A}$ in the anomalous regime is at $\theta=\frac{\pi}{2}$ for all values of $\theta_i$, where the matrix is nilpotent, both eigenvalues are $0$. 

The Henrici departure from normality of $\hat{A}'$ reaches its maximum when the matrix is nilpotent, at $\theta=\arctan{\left(-\sqrt{2}\cot{2\theta_i}\right)}$ which is in the anomalous regime when $\frac{\pi}{4}\leq\theta_i\leq\frac{\pi}{2}$.

The anomalous weak value reaches its maximum at $\theta=\frac{1}{2}\arctan{\left(-\sqrt{2}\cot{2\theta_i}\right)}$.

Here, we have explained in detail how the value of $\theta$ in which the weak value is maximum is at the average between the one in which the Henrici departure from normality of $\hat{A}$ and $\hat{A}'$ are maximum for a specific case. However, this is valid for any two-dimensional quantum system.  
\section{Study of the derivative of the normalized Henrici departure from normality and the numerator of the weak value}\label{appendix:derivative}
The numerator of the weak value of the operator defined in Eq.~\ref{eq:varying_operator} is, assuming $\xi_i=\xi_f=0$, 
\begin{equation}
    O_w=\sqrt{\left(\cos{\theta}\cos{\left(\theta_f+\theta_i\right)}+\cos{\phi}\sin{\theta}\sin{\left(\theta_f+\theta_i\right)}\right)^2+\sin^2{\theta}\sin^2{\left(\theta_f-\theta_i\right)}\sin^2{\phi}}.
\end{equation}
The derivative with respect to $\theta$ is
\begin{equation}
    \frac{dO_w}{d\theta}=\frac{-\sin{2\theta}\left(\cos{\left[2\left(\theta_f-\theta_i\right)\right]}+3\cos{\left[2\left(\theta_f+\theta_i\right)\right]}-2\cos{2\phi}\sin{2\theta_f}\sin{2\theta_i}\right)+4\cos{2\theta}\cos{\phi}\sin{\left[2\left(\theta_f+\theta_i\right)\right]}}{8\sqrt{\left(\cos{\theta}\cos{\left(\theta_f+\theta_i\right)}+\cos{\phi}\sin{\theta}\sin{\left(\theta_f+\theta_i\right)}\right)^2+\sin^2{\theta}\sin^2{\left(\theta_f-\theta_i\right)}\sin^2{\phi}}}.
\end{equation}
The Henrici departure of $\hat{A}$ is
\begin{equation}
    df_n\left(\hat{A}\right)=\frac{1}{2}\sqrt{\left(3+2\cos{4\theta_f}\cos^2{\phi}-\cos{2\phi}\right)\sin^2{\theta}+4\cos^2{\theta}\sin^2{2\theta_f}-2\cos{\phi}\sin{2\theta}\sin{4\theta_f}}\ .
\end{equation}
The derivative of the normalized Henrici departure from normality of $\hat{A}$ is
\begin{equation}
    \frac{df_n\left(\hat{A}\right)}{d\theta}=\frac{-4\cos{2\theta}\cos{\phi}\sin{4\theta_f}+\sin{2\theta}\left(\cos{4\theta_f}\left(3+\cos{2\phi}\right)+2\sin^2{\phi}\right)}{4\sqrt{\left(3+2\cos{4\theta_f}\cos^2{\phi}-\cos{2\phi}\right)\sin^2{\theta}+4\cos^2{\theta}\sin^2{2\theta_f}-2\cos{\phi}\sin{2\theta}\sin{4\theta_f}}}.
\end{equation}
The Henrici departure of $\hat{A}'$ is
\begin{equation}
    df_n\left(\hat{A}'\right)=\frac{1}{2}\sqrt{\left(3+2\cos{4\theta_i}\cos^2{\phi}-\cos{2\phi}\right)\sin^2{\theta}+4\cos^2{\theta}\sin^2{2\theta_i}-2\cos{\phi}\sin{2\theta}\sin{4\theta_i}}\ .
\end{equation}
The derivative of the normalized Henrici departure from normality of $\hat{A}'$ is
\begin{equation}
    \frac{df_n\left(\hat{A}'\right)}{d\theta}=\frac{-4\cos{2\theta}\cos{\phi}\sin{4\theta_i}+\sin{2\theta}\left(\cos{4\theta_i}\left(3+\cos{2\phi}\right)+2\sin^2{\phi}\right)}{4\sqrt{\left(3+2\cos{4\theta_i}\cos^2{\phi}-\cos{2\phi}\right)\sin^2{\theta}+4\cos^2{\theta}\sin^2{2\theta_i}-2\cos{\phi}\sin{2\theta}\sin{4\theta_i}}}.
\end{equation}

\section{Pauli matrices}\label{appendix:Pauli_matrices}
The Pauli matrices are:
\begin{eqnarray}
    \hat{\sigma}_x=\begin{pmatrix}
        0 & 1 \\
        1 & 0
    \end{pmatrix}\hspace{1 cm}
    \hat{\sigma}_y=\begin{pmatrix}
        0 & -i \\
        i & 0
    \end{pmatrix}\hspace{1 cm}
        \hat{\sigma}_z=\begin{pmatrix}
        1 & 0 \\
        0 & -1
    \end{pmatrix}
\end{eqnarray}
\section{Gell-Mann matrices}\label{appendix:gell_mann}
 The Gell-Mann matrices are: 
 \begin{eqnarray}
    \hat{\lambda}_1&=&\begin{pmatrix}
        0 & 1 & 0\\
        1 & 0 & 0 \\
        0 & 0 & 0
    \end{pmatrix}\hspace{1 cm}
        \hat{\lambda}_2=\begin{pmatrix}
        0 & -i & 0\\
        i & 0 & 0 \\
        0 & 0 & 0
    \end{pmatrix}\hspace{1 cm}
     \hat{\lambda}_3=\begin{pmatrix}
        1 & 0 & 0\\
        0 & -1 & 0 \\
        0 & 0 & 0
    \end{pmatrix}
    \\
       \hat{\lambda}_4&=&\begin{pmatrix}
        0 & 0 & 1\\
        0 & 0 & 0 \\
        1 & 0 & 0
    \end{pmatrix}\hspace{1 cm}
           \hat{\lambda}_5=\begin{pmatrix}
        0 & 0 & -i\\
        0 & 0 & 0 \\
        i & 0 & 0
    \end{pmatrix}
    \\
    \hat{\lambda}_6&=&\begin{pmatrix}
        0 & 0 & 0\\
        0 & 0 & 1 \\
        0 & 1 & 0
    \end{pmatrix}\hspace{1 cm}
      \hat{\lambda}_7=\begin{pmatrix}
        0 & 0 & 0\\
        0 & 0 & -i \\
        0 & i & 0
    \end{pmatrix}\hspace{1 cm}
    \hat{\lambda}_8=\frac{1}{\sqrt{3}}\begin{pmatrix}
        1 & 0 & 0\\
        0 & 1 & 0 \\
        0 & 0 & -2
    \end{pmatrix}\hspace{1 cm}
    \end{eqnarray}

\bibliographystyle{unsrt}
\bibliography{biblio}

\end{document}